\documentclass{aa}
\usepackage{graphicx}
\usepackage{epstopdf}
\usepackage{longtable}
\usepackage{txfonts}
\usepackage{natbib}
\usepackage{color}
\bibpunct{(}{)}{;}{a}{}{,}
\newcommand{\ms}{$M_{\odot}$}

\usepackage{amstext}

\begin{document}

   \title{Spectral and spatial imaging of the Be+sdO binary \object{$\varphi$ Persei}
   \thanks{Based on observations with MIRC-6T and VEGA-4T instruments on the CHARA Array}}

   \author{D. Mourard\inst{1}
          \and
          J. D. Monnier\inst{2}
          \and
          A. Meilland\inst{1}
          \and
          D. Gies\inst{3}
          \and
          F. Millour\inst{1}
          \and
          M. Benisty\inst{4}
          \and
          X. Che\inst{2}
          \and
          E. D. Grundstrom\inst{5}
          \and
          R. Ligi\inst{1}
          \and
          G. Schaefer\inst{6}
          \and
          F. Baron\inst{2,3} \and S. Kraus\inst{7} \and M. Zhao\inst{8} \and E. Pedretti\inst{9}
          \and
          P. Berio\inst{1} \and J.M. Clausse\inst{1} \and N. Nardetto\inst{1} \and K. Perraut\inst{4} \and A. Spang\inst{1} \and P. Stee\inst{1} \and I. Tallon-Bosc\inst{10}
          \and
          H. McAlister\inst{3,6} \and T. ten~Brummelaar\inst{6} \and S.T. Ridgway\inst{11} \and J. Sturmann\inst{6} \and L. Sturmann\inst{6} \and N. Turner\inst{6} \and C. Farrington\inst{6}}

\institute{
Laboratoire Lagrange, UMR 7293 UNS-CNRS-OCA, Boulevard de l'Observatoire, B.P. 4229 F, 06304 Nice Cedex 4, France
\email{denis.mourard@oca.eu}
         \and
         Department of Astronomy, University of Michigan, Ann Arbor, MI 48109, USA
                 \and
             CHARA, Georgia State University, P.O. Box 3969, Atlanta, GA 30302-3969, USA
        \and  Univ. Grenoble Alpes, IPAG, F-38000 Grenoble, France, CNRS, IPAG, F-38000 Grenoble, France
     \and
    Physics and Astronomy Department,  Vanderbilt University, 6301 Stevenson Center, Nashville, TN 37235, USA
        \and
             CHARA Array, Mount Wilson Observatory, Mount Wilson, CA 91023, USA
         \and University of Exeter, School of Physics, Stocker Road, Exeter EX4 4QL, UK
        \and Department of Astronomy and Astrophysics, Pennsylvania State University, University Park, PA 16802, USA
        \and Heriot-Watt University, Edinburgh, Scotland EH4 4AS, UK
        \and
        Universit\'{e} de Lyon, 69003 Lyon, France; Universit\'{e} Lyon 1, Observatoire de Lyon, 9 avenue Charles Andr\'{e}, 69230 Saint Genis Laval; CNRS, UMR 5574, Centre de Recherche Astrophysique de Lyon; Ecole Normale Sup\'{e}rieure, 69007 Lyon, France
\and
        National Optical Astronomy Observatory, PO Box 26732, Tucson, AZ 85726, USA}

   \date{;}

  \abstract
   {}
   {The rapidly rotating Be star \object{$\varphi$ Persei} was spun up by mass and angular momentum transfer from a now stripped-down, hot subdwarf companion.  Here we present the first high angular resolution images of \object{$\varphi$ Persei} made possible by new capabilities in long-baseline interferometry at near-IR and visible wavelengths.  We analyzed these images to search for the companion, to determine the binary orbit, stellar masses, and fluxes, and to examine the geometrical and kinematical properties of the outflowing disk surrounding the Be star.}
   {We observed \object{$\varphi$ Persei} with the MIRC and VEGA instruments of the CHARA Array. MIRC was operated in six-telescope mode, whereas VEGA was configured in four-telescope mode with a change of quadruplets of telescopes during two nights to improve the $(u,v)$ plane coverage. Additional MIRC-only observations were performed to track the orbital motion of the companion, and these were fit together with new and existing radial velocity measurements of both stars to derive the complete orbital elements and distance.  We also used the MIRC data to reconstruct an image of the Be disk in the near-IR $H$-band.  VEGA visible broadband and spectro-interferometric H$\alpha$ observations were fit with analytical models for the Be star and disk, and image reconstruction was performed on the spectrally resolved H$\alpha$ emission line data.}
   {The hot subdwarf companion is clearly detected in the near-IR data at each epoch of observation with a flux contribution of 1.5\% in the $H$ band, and restricted fits indicate that its flux contribution rises to 3.3\% in the visible. A new binary orbital solution is determined by combining the astrometric and radial velocity measurements. The derived stellar masses are $9.6 \pm 0.3 M_\odot$ and $1.2 \pm 0.2 M_\odot$ for the Be primary and subdwarf secondary, respectively.  The inferred distance ($186 \pm 3$~pc), kinematical properties, and evolutionary state are consistent with membership of \object{$\varphi$ Persei} in the \object{$\alpha$~Per} cluster. From the cluster age we deduce significant constraints on the initial masses and evolutionary mass transfer processes that transformed the \object{$\varphi$ Persei} binary system. The interferometric data place strong constraints on the Be disk elongation, orientation, and kinematics, and the disk angular momentum vector is coaligned with and has the same sense of rotation as the orbital angular momentum vector. The VEGA visible continuum data indicate an elongated shape for the Be star itself, due to the combined effects of rapid rotation, partial obscuration of the photosphere by the circumstellar disk, and flux from the bright inner disk.}
   {}


   \maketitle

\section{Introduction}
For about 30 years, Be stars have been favorite targets of optical
interferometers \citep{thom,gcasNature}, and they are important objects to study the role of rotation in the link between a star and its close environment. Be stars are rapidly rotating, main-sequence B-stars that eject gas into circumstellar disks \citep{Rivinius2013}.  The ``e'' suffix in their spectral classification refers to the presence of Balmer emission lines that form in their outflowing disks. Such emission appears in B-stars with equatorial velocities above approximately $75\%$ of the critical rate, and
processes related to nonradial pulsation and/or small-scale magnetic fields
probably aid mass loss into the disk \citep{Rivinius2013}. In some circumstances, these processes may produce high-energy X-rays. The Be star \object{$\gamma$~Cassiopeiae}, for example, exhibits thermal X-ray emission.
In a recent series of two papers on \object{$\gamma$~Cas},  \citet{smith} and \citet{stee} have reviewed the most remarkable features of this star:
X-ray activity, magnetic field, critical rotation, Keplerian rotation, disk
elongation, and binarity.  These papers have also shown once more that a
multiwavelength and multitechnique approach is key to making progress in
understanding the physics of these complex systems.

The origin of the rapid spin of Be stars is unknown.  Most Be stars appear to be somewhat evolved objects \citep{McSwain2005,Zorec2005}, so that their fast spin represents processes that occur well into their main-sequence lives.  One possibility is that B-stars spin up as they conclude core H-burning due to the redistribution of internal angular momentum \citep{Ekstrom2008,Granada2013,Granada2014}. A second possibility is that Be stars were spun up by mass and angular momentum transfer in a binary \citep{Pols1991,deMink2013}. In this case, the companion would lose most of its envelope and would appear as a hot, stripped-down He-star remnant. The Be binary star \object{$\varphi$ Persei} represents the first detection of a Be star with a hot, faint companion \citep{Poeckert1981,Thaller1995,Gies1998}, and it is one of three such systems known at present \citep{Peters2013}. Clearly it is important to study such Be binaries to determine the physical properties of the stars at the conclusion of their transformative interaction.

This paper is an analysis of the results obtained on \object{$\varphi$ Persei} thanks to a combined visible and infrared interferometric campaign with the instruments VEGA \citep{vega} and MIRC \citep{mirc} on the CHARA Array \citep{chara} and is complemented with a new analysis of a collection of radial velocity measurements. In Sect.~\ref{phiper} we summarize the known facts on \object{$\varphi$ Persei} relevant to this study, and we compile radial velocity measurements in Sect.~\ref{sec:rv}. The journal of interferometric observations is presented in Sect.~\ref{data}, and MIRC data are presented in Sect.~\ref{sec:MIRC} along with a global astrometric and radial velocity analysis. The VEGA data and analysis are presented in Sect.~\ref{sec:Vega}. The scientific implications of our results are presented and discussed in Sect.~\ref{discussion}, and we conclude in Sect.~\ref{conclusion}.

\section{Main characteristics of \object{$\varphi$ Persei}}
\label{phiper}
The source \object{$\varphi$ Persei} (HD10516, HIP8068) is a bright star ($m_V=4.09$) in the northern sky that is classified as B1.5~V:e-shell \citep{Slettebak1982}. The {\it Hipparcos} parallax \citep{leeuw2007b} is $\pi=4.54\pm0.2$ millisecond of arc (mas), which gives a distance $d=220\pm10$~pc. Using the interstellar Ca II distance scale, \citet{Megier2009} estimate a distance of $d=195\pm32$~pc.

\citet{Campbell1895} published a list of 32 stars presenting both bright and dark hydrogen lines and addressed the question of explaining these interesting and varying features. \object{$\varphi$ Persei} appears in this list, and about 120 years later, we are still considering it as a puzzle. Until 1960, most of the papers published on \object{$\varphi$ Persei} after this discovery were dedicated to samples of stars and statistical studies. More detailed work followed this period, and \citet{Slettebak1966} (confirmed later on by \citealt{Abt2002}) published a measurement of the projected rotational velocity of $V \sin i=410-450$ km~s$^{-1}$, which is estimated to be around $85\%$ of the equatorial breakup velocity.

\citet{Poeckert1981} presented the first evidence of a companion in \object{$\varphi$ Persei} and identified it as a helium star. Based on the velocity curves, he estimated masses of $M_p=21$~\ms~and $M_s=3.4$~\ms. The primary is the Be star, whereas the secondary is the hot companion with an effective temperature estimated to be around 50000K in order to ionize helium. The photospheric spectrum of the hot companion was first detected in spectra from the {\it International Ultraviolet Explorer} satellite by \citet{Thaller1995}, who showed that the spectrum appears similar to that of the sdO6 subdwarf star HD~49798. \citet{Bozic1995} presented a comprehensive review of the available photometric and radial velocity data, and they proposed an orbital solution with a period $P=126.6731$ days, an epoch of Be star superior conjunction $T_{sc}={\rm HJD}2435046.73$ (HJD = heliocentric Julian Date), and revised masses: $M_p$ between 16 and 22~\ms~and $M_s$ between 1.7 and 2.2~\ms. They classified the primary star as B0.5e. The companion was finally confirmed by \citet{Gies1998} thanks to high-resolution UV spectroscopy with the {\it Hubble Space Telescope}. The subdwarf-to-Be flux ratio is $0.165\pm0.006$ (resp.\ $0.154\pm0.009$) for the 1374\AA~region (resp.\ 1647\AA). Moreover, they derived a double-lined solution for the radial velocity curve that yields masses of $M_p=9.3\pm0.3$~\ms~and $M_s=1.14\pm0.04$~\ms. The lower mass estimates with respect to the results from \citet{Bozic1995} are explained by a much smaller secondary semiamplitude in the RV measurements, which leads to a smaller projected semimajor axis and thus lower masses. The subdwarf effective temperature is $T_{\rm eff}=53000\pm3000K$ and the surface gravity is $\log g=4.2\pm0.1$.

Through the study of the helium lines, \citet{Stefl2000} argued for an origin of the emission in the outer parts of the disk surrounding the primary star. Their observations agree with a scenario where the outer parts of an axisymmetric disk are illuminated by the radiation of the secondary. They also favor inhomogeneities in the global density pattern of the inner regions as evidenced by the emission line asymmetry. The model of the Fe~II $5317$\AA~ and He~I $6678$\AA~ and $5876$\AA~ emission lines of \object{$\varphi$ Persei} proposed by \citet{Hummel2001} yields the size and shape of the excitation region in the circumprimary disk. The Fe~II emission originates within 12 stellar radii in an axisymmetric disk around the primary, whereas the He~I emission is best fit by a disk sector with a radius of 15 stellar radii and opening angle of $\simeq120^{\circ}$ facing the secondary \citep{Hummel2003}. They also confirmed that the complex structure of the emission line profiles is due to an external illumination.

Polarimetric studies of \object{$\varphi$ Persei} were made in the late 1990s through the work of \citet{Clarke1998} and \citet{Ghosh1999}. A wavelength dependence of the polarimetric fluctuations was detected and associated with events deep inside the binary system, the amount of scattered radiation being primarily controlled by the opacity of the primary star's disk. \citet{Ghosh1999} measured an intrinsic polarization of \object{$\varphi$ Persei} equal to $1.57\pm0.11\%$ (resp. $1.36\pm0.10\%$, $1.19\pm0.09\%$ and $0.98\pm0.08\%$) in the B (resp., V, R and I) band. The position angle (measured east from the north celestial pole) of the scattering disk normal axis is $PA=27\pm2^{\circ}$.

Optical interferometric observations of \object{$\varphi$ Persei} (see Table~\ref{tab:interfero}) were first reported in 1997 with the MarkIII interferometer \citep{Quirrenbach1997}. Using a wide spectral band of $54$\AA~ centered on H$\alpha$, they fitted their data with a model of a Gaussian disk plus a stellar uniform disk of $0.39$~mas contributing $23\%$ of the flux in the spectral band. Using the NPOI interferometer with an H$\alpha$ filter of width $39$\AA$\pm3$\AA~, \citet{Tycner2006} used the same model as \citet{Quirrenbach1997} and obtained similar results. The authors also derived an inclination angle $i\geq74^{\circ}$. Finally, \citet{Gies2007}, using the CHARA Array in the $K'$-band, fitted a model with a Gaussian elliptical disk plus a central source and a secondary object accounting for $6\%$ of the flux ($\Delta K'=2.9$ mag). The flux contribution from the central source is estimated as $53.8\pm1.5\%$.

\begin{table}[h]
\centering
\caption{Interferometric estimates of disk parameters. References are (1) \citet{Quirrenbach1997}, (2) \citet{Tycner2006}, (3) \citet{Gies2007}, (4) \citet{Touhami2013}, (5) this paper. Columns 1 and 2 list the ratio of the minor to major axis $r$ and its uncertainty, Columns 3 and 4 list the disk position angle along the major axis $PA$ and its uncertainty, and Cols. 5 and 6 list the angular FWHM of the disk major axis $\theta_{\rm maj}$ and its uncertainty.}
\label{tab:interfero}
\setlength{\tabcolsep}{4pt}
\begin{tabular}{cccccccc}
\hline
$r$ & $\delta r$ & $PA$ & $\delta PA$ & $\theta_{\rm maj}$ & $\delta \theta_{\rm maj}$ &  Band & Ref.\\
\hline
0.46  & 0.04    &    -62   & 5   & 2.67 & 0.20  & H$\alpha$ & (1)\\
0.27  & 0.01    &    -61   & 1   & 2.89 & 0.09  & H$\alpha$ & (2)\\
0.00  & 0.22    &    -44   & 3   & 2.30 & 0.08  & $K'$      & (3)\\
0.10  & ....    &    -44   & 4   & 2.44 & 0.22  & $K'$      & (4)\\
0.15  & 0.09    &    -64   & 3   & 1.44 & 0.41  & $H$       & (5)\\
0.37  & 0.03    &    -76   & 1   & 0.48 & 0.05  & $R$       & (5)\\
0.34  & 0.03    &    -69   & 2   & 0.82 & 0.10  & H$\alpha$ & (5)\\
\hline
\end{tabular}
\end{table}

The main features of the system may be summarized as follows: The system is a known spectroscopic binary with a period of almost 127 days. The primary object is a Be star with a very high rotational velocity ($450$ km~s$^{-1}$) that presumably presents a very elongated photosphere seen almost equator-on. As evidenced by the first interferometric measurement and in accordance to its shell classification, it is expected that the inclination angle is at almost $90^\circ$. The companion is a hot subdwarf star. We can estimate from the UV flux a contribution of $8\%$ or less in the $V$ - and $H$-band continuum. The main goals of our study are to attempt a direct detection of the companion in the optical to determine its physical and orbital properties and to continue the study of the disk geometry and kinematics.

\section{Spectroscopic orbital determination from radial velocity measurements}
\label{sec:rv}

A combined analysis of the astrometric and radial velocity orbits
provides the key step to determining the stellar masses and distance.
However, a comprehensive spectroscopic orbit was last determined by
\citet{Bozic1995}, so it is worthwhile to reassess the
spectroscopic orbital elements based on observations that are
more contemporaneous with the interferometric observations.
Here we present new radial velocity measurements for
the Be primary and for the elusive sdO secondary.

\citet{Bozic1995} showed that the wings of the H$\alpha$
emission line that are formed in the inner disk close to the Be star
display radial velocity variations of the Be star.  Thus, we
collected the available H$\alpha$ spectroscopy from the
Be Star Spectra Database\footnote{http://basebe.obspm.fr/basebe/}
\citep{Neiner2011} and from the work of \citet{Grundstrom2007}.
These were supplemented by several spectra we obtained with the
KPNO Coude Feed telescope and one spectrum published by
\citet{Tomasella2010}.  All of these are high S/N spectra with a
resolving power generally better than $R=\lambda/ \triangle\lambda=5000$.
The reduced heliocentric Julian dates (RJD) of observation and origins of
these 79 spectra are listed in Table~\ref{tab:rv}. The H$\alpha$ emission strength changed by a factor of two during these observations (with local maximum strength in 2005 and local minimum strength in 2011), and there were changes visible in the shape of the line core.  However, the shapes of the emission wings remained more or less constant throughout this period, and this suggests that fast-moving gas close to the base of the disk at the photosphere was present throughout. Consequently, measurements of the Doppler shifts from the wing bisector should provide a reasonable estimate of the motion of the gas closest to the Be star itself. \citet{Bozic1995} came to a similar conclusion and found no evidence of systematic deviations in the resulting velocities (which might be present, for example, if high-speed gas streams contributed to the emission flux).

\begin{longtab}
\begin{longtable}{lccl}
\caption{\label{tab:rv}Radial velocity measurements}\\
\hline
Component      & RJD        & Vr (km~s$^{-1}$)         & Source and Observer\\
\hline
Be Primary     & 50023.5987 &   -9.8 $\pm$  1.5 & HST   Gies et al. (1998)\\
Be Primary     & 50084.5151 &    8.7 $\pm$  1.5 & HST   Gies et al. (1998)\\
Be Primary     & 50276.5235 &  -12.8 $\pm$  1.5 & HST   Gies et al. (1998)\\
Be Primary     & 50327.3240 &    4.1 $\pm$  1.5 & HST   Gies et al. (1998)\\
Be Primary     & 50371.3607 &   -0.2 $\pm$  1.5 & HST   Gies et al. (1998)\\
Be Primary     & 50371.4089 &   -0.2 $\pm$  1.5 & HST   Gies et al. (1998)\\
Be Primary     & 50709.6087 &    1.9 $\pm$  1.0 & BeSS  Neiner\\
Be Primary     & 50709.6193 &    2.0 $\pm$  1.0 & BeSS  Neiner\\
Be Primary     & 51056.9586 &  -22.0 $\pm$  1.1 & KPNO  Gies\\
Be Primary     & 51221.3465 &   -0.2 $\pm$  0.8 & INAF  Tomasella et al. (2010)\\
Be Primary     & 51419.9373 &  -11.1 $\pm$  3.8 & KPNO  Gies\\
Be Primary     & 51419.9390 &  -11.3 $\pm$  3.5 & KPNO  Gies\\
Be Primary     & 51420.9717 &  -11.6 $\pm$  5.0 & KPNO  Gies\\
Be Primary     & 51421.9875 &  -14.5 $\pm$  2.6 & KPNO  Gies\\
Be Primary     & 51421.9905 &  -13.1 $\pm$  2.7 & KPNO  Gies\\
Be Primary     & 51430.9768 &  -14.7 $\pm$  4.0 & KPNO  Gies\\
Be Primary     & 51430.9805 &  -14.8 $\pm$  4.2 & KPNO  Gies\\
Be Primary     & 52237.3904 &   -5.2 $\pm$  0.7 & BeSS  Buil\\
Be Primary     & 52241.3562 &  -10.1 $\pm$  0.7 & BeSS  Buil\\
Be Primary     & 52253.3768 &   -4.9 $\pm$  0.7 & BeSS  Buil\\
Be Primary     & 52263.2743 &   -2.3 $\pm$  0.4 & BeSS  Buil\\
Be Primary     & 52284.3159 &  -18.4 $\pm$  0.7 & BeSS  Buil\\
Be Primary     & 52293.3072 &  -18.0 $\pm$  0.6 & BeSS  Buil\\
Be Primary     & 52303.2874 &  -29.4 $\pm$  0.6 & BeSS  Buil\\
Be Primary     & 52521.4100 &   10.7 $\pm$  0.9 & BeSS  Buil\\
Be Primary     & 52854.5309 &   13.4 $\pm$  1.0 & BeSS  Buil\\
Be Primary     & 52855.5299 &   10.3 $\pm$  1.5 & BeSS  Buil\\
Be Primary     & 53290.8068 &   -0.7 $\pm$  1.2 & KPNO  Grundstrom\\
Be Primary     & 53290.8076 &   -0.6 $\pm$  1.4 & KPNO  Grundstrom\\
Be Primary     & 53290.8087 &   -0.0 $\pm$  1.6 & KPNO  Grundstrom\\
Be Primary     & 53291.7942 &   -0.8 $\pm$  1.1 & KPNO  Grundstrom\\
Be Primary     & 53291.7951 &   -0.3 $\pm$  1.3 & KPNO  Grundstrom\\
Be Primary     & 53292.8129 &   -1.6 $\pm$  1.0 & KPNO  Grundstrom\\
Be Primary     & 53292.8140 &   -2.1 $\pm$  1.1 & KPNO  Grundstrom\\
Be Primary     & 53294.7812 &   -0.1 $\pm$  2.2 & KPNO  Grundstrom\\
Be Primary     & 53294.7851 &   -0.1 $\pm$  1.7 & KPNO  Grundstrom\\
Be Primary     & 54019.8264 &   10.2 $\pm$  1.2 & KPNO  Grundstrom\\
Be Primary     & 54020.8271 &    8.1 $\pm$  0.9 & KPNO  Grundstrom\\
Be Primary     & 54021.8123 &    6.6 $\pm$  1.2 & KPNO  Grundstrom\\
Be Primary     & 54022.7978 &    8.3 $\pm$  1.4 & KPNO  Grundstrom\\
Be Primary     & 54023.8106 &    7.0 $\pm$  0.9 & KPNO  Grundstrom\\
Be Primary     & 54331.5338 &  -24.6 $\pm$  2.1 & BeSS  Desnoux\\
Be Primary     & 54358.5030 &  -18.2 $\pm$  0.9 & BeSS  Thizy\\
Be Primary     & 54393.5289 &    5.6 $\pm$  1.2 & BeSS  Thizy\\
Be Primary     & 54513.3741 &   12.7 $\pm$  2.4 & BeSS  Guarro\\
Be Primary     & 54672.6520 &    3.2 $\pm$  1.0 & BeSS  Buil\\
Be Primary     & 54761.7315 &   -0.6 $\pm$  3.2 & KPNO  Grundstrom\\
Be Primary     & 54793.4824 &    8.3 $\pm$  1.1 & BeSS  Guarro\\
Be Primary     & 54866.2563 &  -27.1 $\pm$  2.1 & BeSS  Guarro\\
Be Primary     & 55059.5401 &   -8.5 $\pm$  3.5 & BeSS  Desnoux\\
Be Primary     & 55060.5620 &   -3.6 $\pm$  1.2 & BeSS  Buil\\
Be Primary     & 55067.5022 &   -9.5 $\pm$  1.4 & BeSS  Thizy\\
Be Primary     & 55072.3554 &   11.0 $\pm$  4.3 & BeSS  Terry\\
Be Primary     & 55109.4944 &  -12.6 $\pm$  2.8 & BeSS  Buil\\
Be Primary     & 55117.3864 &  -10.7 $\pm$ 10.7 & BeSS  Garrel\\
Be Primary     & 55119.3849 &  -15.7 $\pm$ 11.4 & BeSS  Garrel\\
Be Primary     & 55203.3741 &  -17.7 $\pm$ 17.1 & BeSS  Garrel\\
Be Primary     & 55398.4875 &   11.1 $\pm$  1.3 & BeSS  Terry\\
Be Primary     & 55417.5052 &    8.0 $\pm$  2.5 & BeSS  Desnoux\\
Be Primary     & 55455.4248 &  -16.4 $\pm$  1.2 & BeSS  Buil\\
Be Primary     & 55534.2922 &   -1.9 $\pm$  1.6 & BeSS  Guarro\\
Be Primary     & 55559.4242 &   -0.3 $\pm$  2.0 & BeSS  Garrel\\
Be Primary     & 55774.5908 &    3.2 $\pm$  2.2 & BeSS  Ubaud\\
Be Primary     & 55777.5094 &   -2.6 $\pm$  0.8 & BeSS  Dubreuil\\
Be Primary     & 55785.5531 &    2.5 $\pm$  0.7 & BeSS  Terry\\
Be Primary     & 55828.3888 &  -17.2 $\pm$  0.6 & BeSS  Terry\\
Be Primary     & 55831.5150 &  -20.7 $\pm$  1.3 & BeSS  Buil\\
Be Primary     & 55857.4198 &  -31.3 $\pm$  0.6 & BeSS  Buil\\
Be Primary     & 55879.6551 &   -9.6 $\pm$  0.6 & BeSS  Graham\\
Be Primary     & 55907.6115 &    8.0 $\pm$  0.6 & BeSS  Graham\\
Be Primary     & 55956.3393 &  -12.3 $\pm$  2.1 & BeSS  Garrel\\
Be Primary     & 56143.6007 &    5.8 $\pm$  1.3 & BeSS  Buil\\
Be Primary     & 56150.4413 &    7.5 $\pm$  2.8 & BeSS  Terry\\
Be Primary     & 56163.4058 &   17.4 $\pm$  2.6 & BeSS  Garrel\\
Be Primary     & 56167.4575 &   18.1 $\pm$  1.8 & BeSS  Terry\\
Be Primary     & 56204.5901 &    0.8 $\pm$  2.2 & BeSS  Favaro\\
Be Primary     & 56206.4024 &   -0.4 $\pm$  0.7 & BeSS  Buil\\
Be Primary     & 56260.5174 &   -4.8 $\pm$  0.5 & BeSS  Graham\\
Be Primary     & 56267.3228 &    2.2 $\pm$  1.2 & BeSS  Garrel\\
Be Primary     & 56270.4680 &    4.7 $\pm$  2.0 & BeSS  Guarro\\
Be Primary     & 56291.3623 &   20.6 $\pm$  0.5 & BeSS  Ubaud\\
Be Primary     & 56301.3892 &   19.6 $\pm$  3.4 & BeSS  Guarro\\
Be Primary     & 56490.5331 &   -4.6 $\pm$  1.3 & BeSS  Ubaud\\
Be Primary     & 56628.5588 &    4.6 $\pm$  1.6 & BeSS  Graham\\
Be Primary     & 56665.6322 &   28.1 $\pm$  1.0 & BeSS  Sawicki\\
sdO Secondary  & 43763.4936 & -114.3 $\pm$  3.5 & IUE   \\
sdO Secondary  & 43837.9529 &   62.5 $\pm$  3.6 & IUE   \\
sdO Secondary  & 43839.8268 &   48.4 $\pm$  2.6 & IUE   \\
sdO Secondary  & 43844.0271 &   35.0 $\pm$  3.2 & IUE   \\
sdO Secondary  & 44151.5341 &  -66.7 $\pm$  4.3 & IUE   \\
sdO Secondary  & 44167.3172 &   -8.6 $\pm$  3.8 & IUE   \\
sdO Secondary  & 44174.0036 &    7.6 $\pm$  3.8 & IUE   \\
sdO Secondary  & 44328.2958 &   81.8 $\pm$  4.5 & IUE   \\
sdO Secondary  & 44332.4182 &   86.0 $\pm$  4.0 & IUE   \\
sdO Secondary  & 44485.0176 &  -13.3 $\pm$  2.7 & IUE   \\
sdO Secondary  & 44486.1504 &  -16.9 $\pm$  2.7 & IUE   \\
sdO Secondary  & 44511.1058 &  -96.3 $\pm$  2.6 & IUE   \\
sdO Secondary  & 44661.1698 &  -82.7 $\pm$  2.6 & IUE   \\
sdO Secondary  & 44824.9353 &   66.5 $\pm$  2.6 & IUE   \\
sdO Secondary  & 45237.9853 &   28.6 $\pm$  3.2 & IUE   \\
sdO Secondary  & 45238.0244 &   26.5 $\pm$  4.4 & IUE   \\
sdO Secondary  & 50023.5987 &   78.2 $\pm$  1.1 & HST   Gies et al. (1998)\\
sdO Secondary  & 50084.5151 &  -77.6 $\pm$  1.1 & HST   Gies et al. (1998)\\
sdO Secondary  & 50276.5235 &   75.5 $\pm$  1.1 & HST   Gies et al. (1998)\\
sdO Secondary  & 50327.3240 &  -54.1 $\pm$  1.1 & HST   Gies et al. (1998)\\
sdO Secondary  & 50371.3607 &  -24.2 $\pm$  1.1 & HST   Gies et al. (1998)\\
sdO Secondary  & 50371.4089 &  -23.0 $\pm$  1.1 & HST   Gies et al. (1998)\\
\hline
\end{longtable}
\end{longtab}

We measured the H$\alpha$ wing velocity by determining the emission line
bisector position using the method described by \citet{Shafter1986}.
The radial velocity is found by cross-correlating the observed profile
with a pair of oppositely signed Gaussian functions offset to positions
of $\pm 250$ km~s$^{-1}$ from line center and then measuring the
zero-crossing position of the resulting cross-correlation function.
The uncertainties in the resulting measurements were estimated using
the semi-empirical expression from \citet{Grundstrom2007}, but we
caution that these estimates do not account for long-term changes
in the emission profile that also influence the wing measurements.
In most cases we were able to also measure accurate positions of the
nearby atmospheric telluric lines, and we used these to make small
corrections to the wavelength calibration in the topocentric frame.
The resulting radial velocity measurements appear in Col. 3 of Table~\ref{tab:rv}.
For convenience, we also include a set of six
radial velocities measured by \citet{Gies1998} for the Be star
photospheric lines in far-ultraviolet spectra obtained with the
{\it Hubble Space Telescope}.

We made a single-lined spectroscopic orbital determination from these radial
velocities alone using the program SBCM described by \citet{Morbey1974}.
We began making circular fits with the period derived by \citet{Bozic1995},
$P=126.6731\pm0.0071$~d, and we derived an orbital semiamplitude of
$K_a = 11.4 \pm 1.0$ km~s$^{-1}$.  This agrees within the uncertainties
with the value reported by \citet{Gies1998}, $K_a = 10.0 \pm 0.8$ km~s$^{-1}$.

The measurement of radial velocities for the sdO companion is much more
difficult. There are several features in the optical spectrum that display
the antiphase Doppler shifts expected for the companion, but these are
usually associated with gas near the companion star.  These include weak
He~II $4686$\AA~ emission that may form in a disk close to the companion
\citep{Poeckert1981}, He~I $4026$\AA~, $4471$\AA~ ``shell'' lines
formed farther away \citep{Poeckert1981}, and the He~I $6678$\AA~
emission line that probably originates in the outer disk of the Be star
facing the hot subdwarf \citep{Stefl2000,Hummel2001}. There are only six
{\it HST} measurements of the photospheric lines of the companion itself that were made in the far-ultraviolet spectrum by \citet{Gies1998}, and we include these in Table~\ref{tab:rv}. There are also 16 high-resolution FUV spectra of \object{$\varphi$ Persei} available in the archive of the {\it International Ultraviolet Explorer Satellite} \citep{Thaller1995}, and we successfully detected the spectral signature of the companion in each of these using the cross-correlation methods described by \citet{Peters2013}.
We measured radial velocities for the companion using the same model template
spectrum adopted in the work of \citet{Gies1998}, and our derived radial
velocities appear at the bottom of Table~\ref{tab:rv}. The uncertainties
associated with these velocities are probably larger than reported in the table. Indeed, since they are too faint, the interstellar lines in the spectrum of \object{$\varphi$ Persei} cannot be used to correct for zero-point shifts of the star position with respect to the slit in the large-aperture observations \citep{Peters2013}.

We made a single-lined spectroscopic orbital solution for the secondary again using the orbital period from \citet{Bozic1995}, and we obtained a semiamplitude of $K_b = 83.6 \pm 2.1$ km~s$^{-1}$ that agrees within errors with the determination from the {\it HST} spectra of $K_b = 81.2 \pm 0.6$ km~s$^{-1}$ \citep{Gies1998}. Elliptical fits provided no improvement
in the residuals. Thus, the {\it IUE} measurements tend to confirm the results from the {\it HST} spectra.  On the other hand, because the measurement errors are larger for the {\it IUE} velocities, it is not clear whether or not
including them improves our estimate of the semiamplitude $K_b$.
In the following section, we restrict our considerations to the
{\it HST} velocities of the companion.  Note that the FUV derived estimate of the
secondary semiamplitude is somewhat lower than that found from optical studies
($K_b = 105 \pm 7$ km~s$^{-1}$ from He~II $4686$\AA~, \citealt{Poeckert1981};
 $K_b  \approx 90$ km~s$^{-1}$ from He~I $4026$\AA~, $4471$\AA~, \citealt{Poeckert1981};  $K_b = 101 \pm 6$ km~s$^{-1}$ from He~I $6678$\AA~, \citealt{Bozic1995})
presumably because the latter are influenced by gas motions and a location of
origin that differ from that of the hot companion. The He~I $6678$\AA~ emission, for example, probably originates in the outer disk facing the companion at a position closer to the Be star and thus with a higher Keplerian orbital speed than that of the companion.

\section{Interferometric observations and data analysis principles}
\label{data}
\begin{figure}[h]
\center
 \includegraphics[width=6cm,height=6cm]{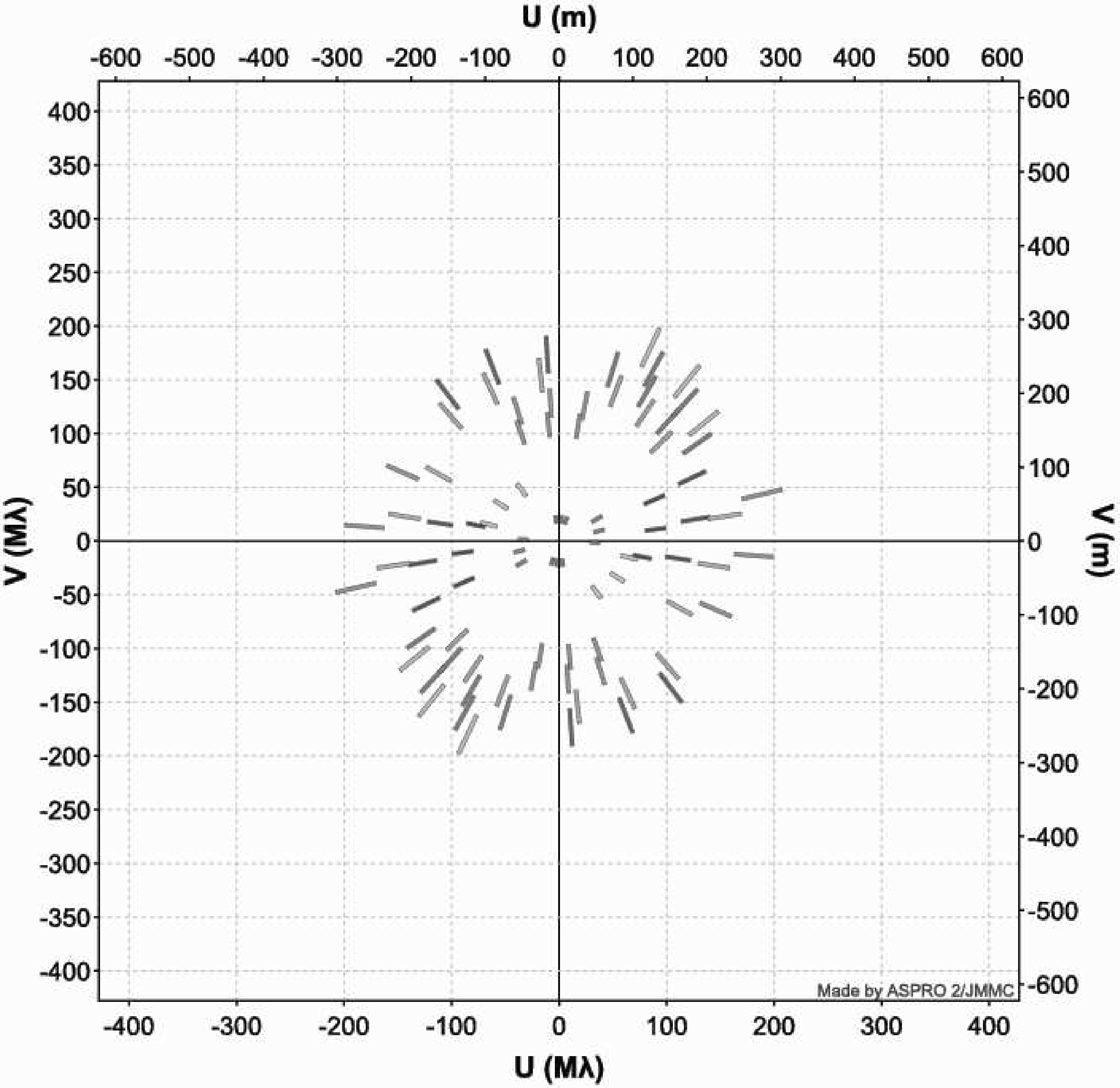}\\
 \includegraphics[width=6cm,height=6cm]{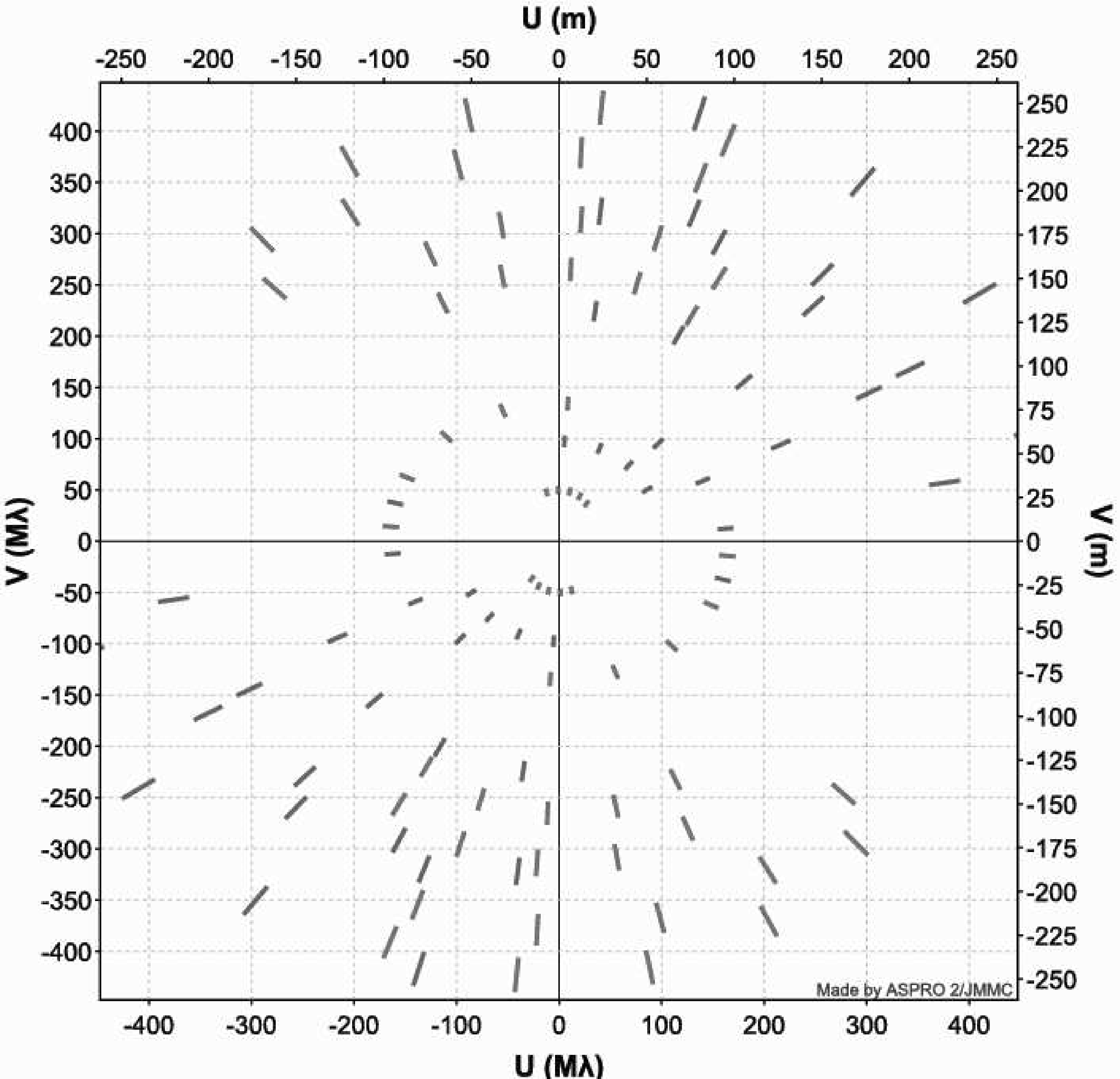}
  \caption{Top: $(u,v)$ plane coverage for MIRC. Bottom: $(u,v)$ plane coverage for VEGA. The two figures use the same spatial frequency scale. }
  \label{fig:planuv}
\end{figure}

Interferometric observations were carried out on the CHARA Array \citep{chara}, located at Mount Wilson Observatory. It consists of six 1~m telescopes sending their beams, after compression, through vacuum pipes to the recombining lab, through the fixed and movable delay lines with the correction of longitudinal dispersion, and the beams are re-arranged before feeding the instruments. For the purpose of this work we have used the MIRC instrument \citep{mirc2004,mirc} operated in the $H$ band using all six CHARA telescopes \citep{mirc2012} and the VEGA spectrograph used in the 4-T mode \citep{vega2}. The spectral configurations used with the different instruments are summarized in Table~\ref{tab:spectro}. The different configurations we used during these observations are presented in Table~\ref{tab:obs}. The corresponding $(u,v)$ coverage, generated with the \texttt{Aspro2} service\footnote{Available at http://www.jmmc.fr/aspro}, is presented in Fig.~\ref{fig:planuv}.

\begin{table}[h]
\centering
\caption{Spectral configurations used with VEGA and MIRC for the \object{$\varphi$ Persei} observations. Column 2 gives the spectral band, Col. 3 the spectral width of one channel, and Col. 4 the number of channels.}
\label{tab:spectro}
\setlength{\tabcolsep}{4pt}
\begin{tabular}{lccc}
\hline
& $\lambda$~nm & $\Delta\lambda$~nm & Channels\\
\hline
VEGA & [483;498] & 15 & 1\\
& [636;651] & 15 & 1\\
& [648;663] & 15 & 1\\
& [663;678] & 15 & 1\\
\hline
VEGA & [646;666] & 0.1 & 200\\
\hline
MIRC & [1501;1712] & 35 & 6\\
\hline
\end{tabular}
\end{table}

For the VEGA observations, VEGA and MIRC were used simultaneously, MIRC with six telescopes in $H$ band and VEGA with two different four-telescope configurations. MIRC was used as a 6-T group delay sensor, and corrections were sent directly to the main delay lines. An initial cophasing of the two instruments was made using a reference target thanks to the fine adjustments of the MIRC feeding optics and corresponding corrections on the main delay lines. The observing sequence was then 1) MIRC finds and stabilizes the fringes, 2) MIRC and VEGA start recording data, 3) at the end of the VEGA sequence, MIRC does its shutter and photometry sequence.

\begin{table}[h]
\centering
\caption{\object{$\varphi$ Persei} observations with VEGA and MIRC on the CHARA Array. Column~1 gives the UT date, Column~2 the reduced heliocentric Julian date RJD, Col.~3 the orbital phase of the companion according to our newly determined ephemeris, Col.~4 the telescope configuration, and Col.~5 the UT time of observation. } \label{tab:obs}
\setlength{\tabcolsep}{4pt}
\begin{tabular}{lcccc}
\hline
Date & RJD & Phase & Telescopes & UT\\
\hline
VEGA\\
\hline
2011Sep28 & 55832.86 & 0.81 & S2E2W1W2 & 08:40\\
2011Oct18 & 55852.77 & 0.97 & S1S2W1W2 & 06:26\\
2011Oct18 & 55852.85 & 0.97 & S1S2W1W2 & 08:31\\
2011Oct18 & 55852.96 & 0.97 & E1E2W1W2 & 10:58\\
2011Oct19 & 55853.72 & 0.98 & S1S2W1W2 & 05:16\\
2011Oct19 & 55853.78 & 0.98 & S1S2W1W2 & 06:39\\
2011Oct19 & 55853.88 & 0.98 & E1E2W1W2 & 09:02\\
2011Oct19 & 55853.93 & 0.98 & E1E2W1W2 & 10:21\\
\hline
MIRC\\
\hline
2011Sep03 & 55807.91 & 0.62 & S1S2E1E2W1W2 & 09:50 \\
2011Sep28 & 55832.86 & 0.81 & S1S2E1E2W1W2 & 08:38\\
2011Oct18 & 55852.87 & 0.97 & S1S2E1E2W1W2 & 08:52 \\
2011Oct19 & 55853.76 & 0.98 & S1S2E1E2W1W2 & 06:14 \\
2012Aug18 & 56158.02 & 0.38 & S1S2E1E2W1W2 & 12:29\\
2012Sep15 & 56185.81 & 0.60 & S1S2E1E2W1W2 & 07:26\\
2012Oct31 & 56231.77 & 0.96 & S1S2E1E2W1W2 & 06:29 \\
2012Nov06 & 56237.61 & 0.01 & S1S2E1E2W1W2 & 02:38\\
2013Oct08 & 56573.86 & 0.66 & S1S2E1E2W1W2 & 08:38 \\
2013Oct21 & 56586.76 & 0.76 & S1S2E1E2W1W2 & 06:14\\
\hline
\end{tabular}
\end{table}

Two different calibrators were used for the VEGA observations: \object{HD3360} and \object{HD25642}. They were identified using the \texttt{SearchCal} service\footnote{Available at http://www.jmmc.fr/searchcal} \citep{SearchcalBright,SearchcalFaint} and the CDS Astronomical Databases SIMBAD and VIZIER\footnote{Available at http://cdsweb.u-strasbg.fr/}. We used the \texttt{SearchCal} spectrophotometric uniform disk angular diameter determinations in the $R$ and $H$ bands, which led to $\theta_{UD-R}=0.284\pm0.020$~mas and $\theta_{UD-H}=0.287\pm0.020$~mas for \object{HD3360} and to $\theta_{UD-R}=0.439\pm0.031$~mas and $\theta_{UD-H}=0.447\pm0.031$~mas for \object{HD25642}.  In addition to these calibrators, MIRC calibration employed the additional calibrators \object{7 And} ($0.65\pm0.03$~mas), \object{$\theta$~Cas} ($0.57\pm0.04$~mas), and \object{HD~33167} ($0.50\pm0.04$~mas).

The data processing followed the standard MIRC \citep{monnier2007,monnier2012} and VEGA \citep{vega,vega2} procedures. Based upon our understanding of the current VEGA limitations \citep{vegaspie2012}, we decided to reject all closure phase measurements with VEGA and to set a conservative lower limit to the absolute uncertainty of the squared visibility measurements at the level of 0.05.

\section{Near-IR image reconstruction and companion detection}
\label{sec:MIRC}
\subsection{MIRC results}
\label{sec:MIRC_results}

\begin{figure*}[!th]
\center
 \includegraphics[width=16cm]{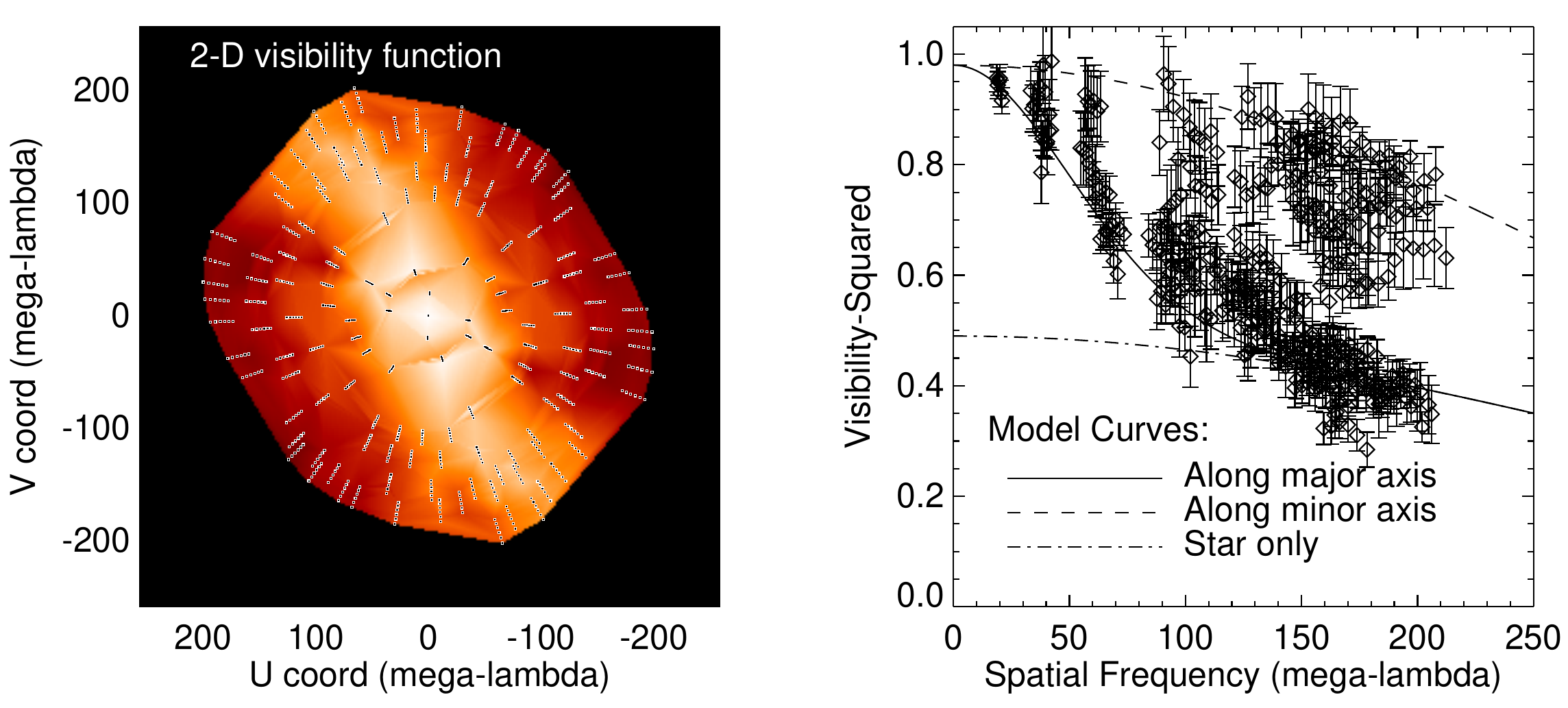}\\
  \caption{(left) Here we show the two-dimension visibility function for \object{$\varphi$ Persei} from MIRC.  We combined all our epochs together and averaged with a 30m smoothing window, maintaining each wavelength channel.  (right) This shows the same averaged data as a function of radial spatial frequency unit.  In addition, we have included model visibilities for position angles along the major and minor axes of the average disk model (see Table~\ref{tab:MIRCmodel}).}
  \label{fig:mircplot}
\end{figure*}

For the purposes of this paper, we wished to extract two major findings from the MIRC data.  First, we sought to directly
detect the companion for the first time to determine the visual orbit of the \object{$\varphi$ Persei} binary system.  Secondly, we aimed to take advantage of the dense and complete $(u,v)$-coverage from MIRC to create an image using aperture synthesis and to constrain the inclination angle of the disk.

\begin{figure}[h]
\center
 \includegraphics[height=9cm,angle=90]{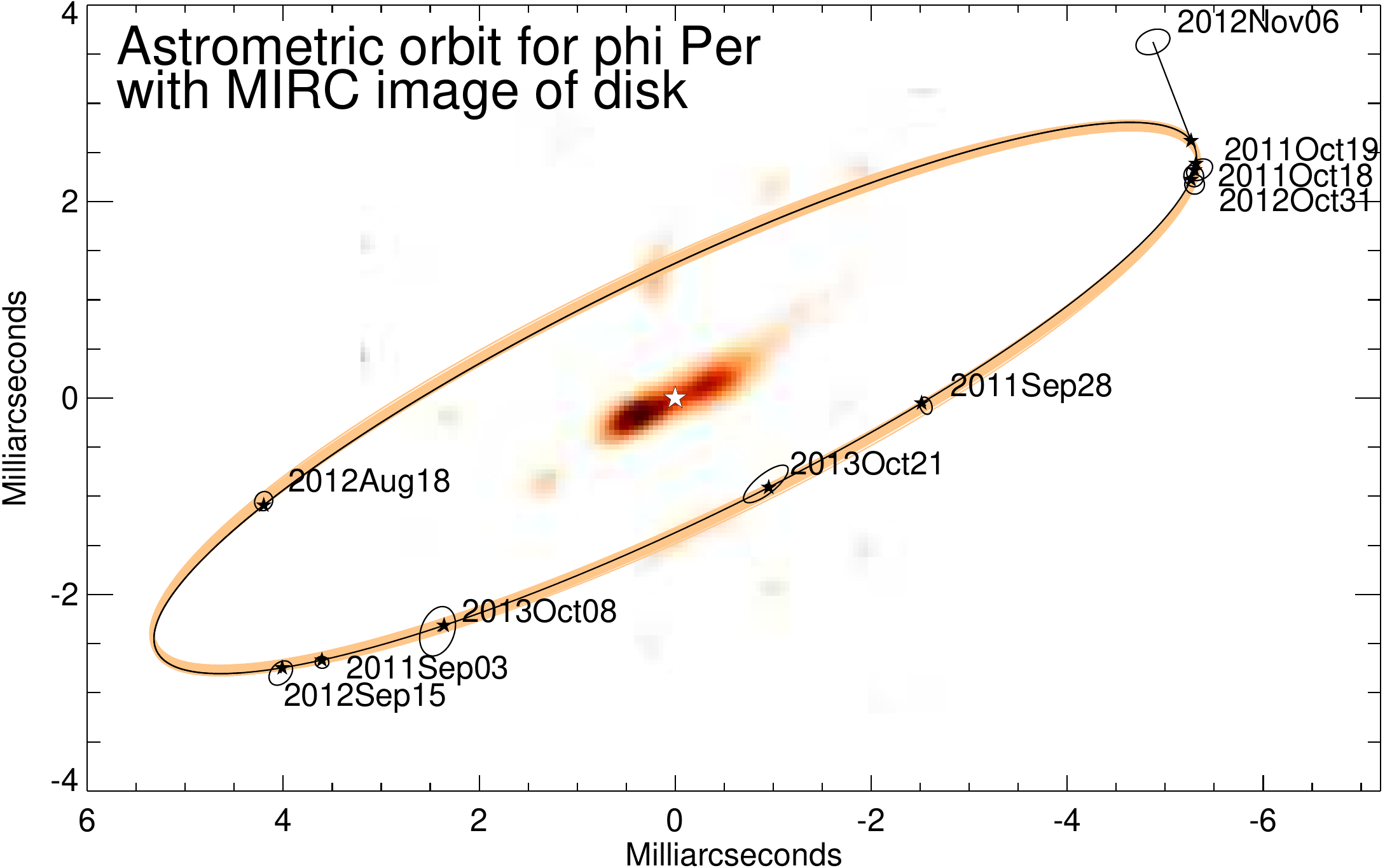}\\
  \caption{Image reconstruction made with MIRC data in the $H$
band. Here we have separately modeled the contributions from the Be star and the companion, and this image represents the emission from the {\em \textup{disk alone}}, contributing about $\sim$29\% of the total $H$-band flux. We have marked the location of the companion at the ten epochs reported here with error ellipses and show the best-fit circular orbit (see text). The colored band shows the results from 100 bootstrap fits to illustrate the errors on our orbital solution.  East is left and north is up.}
  \label{fig:diskMIRC}
\end{figure}

First, to extract the companion location, we fitted a binary model to each of the ten best observing epochs with six-telescope MIRC data.  The model for each epoch has 11 parameters: fraction of $H$-band light from the primary, uniform disk diameter for the primary, fraction of light from the companion, uniform disk diameter for the companion, separation and position angle of the companion compared to the primary, Gaussian model for disk emission with three parameters describing the symmetric component (FWHM of the major axis, FWHM of the minor axis, position angle on the sky of the major axis), and two parameters to describe any asymmetric disk emission \citep[skew angle and skew fraction; see description in][]{monnier2006,che2012}.
In our case, the uniform disk diameter (UDD) of the primary was fixed at 0.3 mas and the UDD of the companion was fixed as 0.056 mas, on the basis of spectrophotometric estimations. We might expect the skewness to change from epoch to epoch \citep[as was found for \object{$\delta$~Sco} by][]{che2012}. This is indeed confirmed by our results at each epoch, but unfortunately, no conclusion could be drawn. Based on our fitting, we report the median and one-sigma range of the $H$-band disk parameters in Table~\ref{tab:MIRCmodel}, while the locations of the companion for all epochs are collected in Table~\ref{tab:companionMIRC}. The typical reduced $\chi^2$ achieved during fitting was $\sim$1.9. A synthetic view of the MIRC data is presented in Fig.~\ref{fig:mircplot}.

We used the MACIM imaging algorithm \citep{macim2006} along with the dark energy regularizer to carry out image reconstruction by combining all the MIRC data together.  Since the primary star contains $\sim$70\% of the total flux, it is important to remove this component explicitly as part of the image reconstruction, as explained in detail in \citet{che2012}.  We accounted for the moving secondary based on an orbital model, although this has only a small effect since the companion only contributes $\sim1.5\%$ of the total light in the system. Figure~\ref{fig:diskMIRC} shows our image reconstruction of the disk alone, along with the locations of the secondary star for each epoch.  The dynamic range appears to be noticeably smaller than most MIRC imaging (background noise is about ten times below peak emission), but this is because the central star has been removed from this image through modeling.  If we were to add back the central star, we would find a dynamic range of $\sim$100 in this image. The reduced $\chi^2$ for this image is $\sim$2.0, only slightly worse than found using disk modeling for each individual epoch.  We expect some time variation between epochs, and so it is not too surprising that the $\chi^2$ is not as low as typically found when imaging with MIRC.

We briefly mention some known limitations of our method and point out future work.  First, the disk model is quite simplistic. As discussed elsewhere \citep[e.g., ][]{Kraus2012, che2012}, the gas disk is expected to have a hole in the center where the star is, and we also expect significant obscuration of the central star at this high-inclination angle.  Furthermore, the imaging suggests a Gaussian disk profile, which is probably not a realistic model.  The asymmetric component of the disk emission changes the photo center of the primary object, thus adding errors to our estimate of the secondary separation vector.  Indeed, it seems the companion location of 2012Nov06 is physically impossible, and our estimate might have been affected by changing asymmetries in the inner disk that were not accurately modeled by our method.

\begin{table}[h]
\centering
\caption{Basic parameters of the $H$-band disk model from MIRC observations. The component fluxes are indicated by $f_a$, $f_d$, and $f_b$
for the Be star, disk, and companion, respectively.}
\label{tab:MIRCmodel}
\setlength{\tabcolsep}{4pt}
\begin{tabular}{c|c}
\hline
Model Parameter & Value \\
\hline
$f_a$ & 0.70$\pm$0.06 \\
$f_d$ & 0.29$\pm$0.06 \\
$f_b$ & 0.015$\pm$0.003  \\
FWHM$_{\rm major}$(mas) & 1.44$\pm$0.41 \\
FWHM$_{\rm minor}$ (mas) &  0.22$\pm$0.12 \\
PA of major axis (deg) &  -64$\pm$3 \\
Disk inclination (deg) & 82$\pm$4 \\
\hline
\end{tabular}
\end{table}

\begin{table}[h]
\centering
\caption{Position of the companion as a function of the RJD as detected in the various MIRC datasets. Separation is given in mas and position angle in degrees. Columns~5--7 give the uncertainty ellipse.} \label{tab:companionMIRC}
\setlength{\tabcolsep}{4pt}
\begin{tabular}{ccccccc}
\hline
RJD & Phase & Separation & $PA$ & $\sigma_{major}$ & $\sigma_{minor}$ & $PA(\sigma)$\\
\hline
55807.91 & 0.62 & 4.50 & 126.7 & 0.07 & 0.06 & 84\\
55832.86 & 0.81 & 2.56 & 268.2 & 0.09 & 0.06 & 18\\
55852.87 & 0.97 & 5.75 & 293.1 & 0.11 & 0.10 & 12\\
55853.76 & 0.98 & 5.83 & 293.5 & 0.14 & 0.10 & 298\\
56158.02 & 0.38 & 4.34 & 104.0 & 0.10 & 0.09 & 330\\
56185.81 & 0.60 & 4.90 & 124.8 & 0.14 & 0.10 & 318\\
56231.77 & 0.96 & 5.73 & 292.3 & 0.10 & 0.10 & 291\\
56237.61 & 0.01 & 6.08 & 306.6 & 0.18 & 0.12 & 292\\
56573.86 & 0.66 & 3.39 & 134.4 & 0.26 & 0.17 & 340\\
56586.76 & 0.76 & 1.27 & 226.6 & 0.28 & 0.10 & 309\\
\hline
\end{tabular}
\end{table}

\subsection{Orbital solution for the companion}
\begin{figure}[h]
\center
\includegraphics[height=9cm,angle=90]{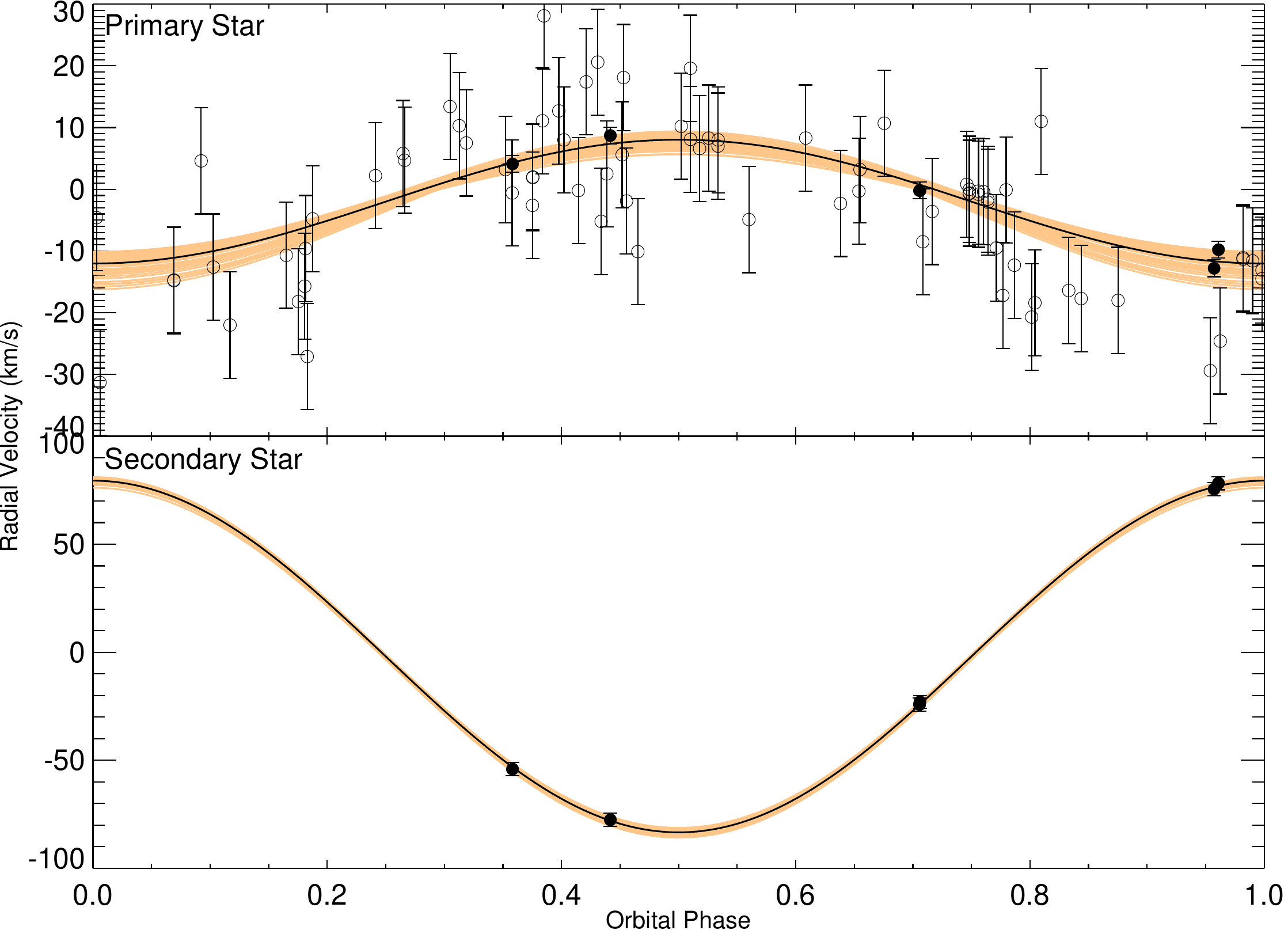}\\
  \caption{Radial velocity measurements of the primary star along with predicted orbit from the joint MIRC$+$RV fit (see Table~\ref{tab:orbit}). Open symbols represent H$\alpha$ wing bisector measurements, while filled symbols represent {\it HST} measurements from \citet{Gies1998}. $1-\sigma$ error bars have been fixed to $\pm 8.6$ km~s$^{-1}$ as explained in the text. The width of the solid line illustrates the errors on our orbital solution estimated from 100 bootstrap fits. The RV is negative at zero phase because we have chosen an astrometric orbital ephemeris of the secondary relative to the primary, which is opposite in sign to the radial velocity convention, where the velocity of the primary star is measured relative to the center-of-mass.}
  \label{fig:vr}
\end{figure}

Based on the radial velocity measurements presented in Sect.~\ref{sec:rv} and
on the companion positions determined in Sect.~\ref{sec:MIRC_results},
we performed a joint astrometric and radial velocity solution of the orbit.
We adopted the period from \citet{Bozic1995}, $P=126.6731\pm0.0071$~d,
in the first combined fit. By combining the epochs of the ascending node of the companion (equal to the velocity maximum for the secondary and the velocity minimum for the Be star primary) from \citet{Bozic1995},  \citet{Gies1998} (119 orbital cycles later), and our new solution (166 orbital cycles later), we then derived improved estimates for the period and epoch. The revised period $P=126.6982\pm0.0035$~d is $3\sigma$ larger than that found by \citet{Bozic1995}, but we use this revised period
and epoch ($T_{\rm RV~min}=RJD~56110.00\pm0.12$) to reference the orbital
phase in the rest of this paper.

We fixed the period to $P=126.6982$ and assumed a circular orbit for our final solution. To assess the errors in the orbital parameters, we carried out a bootstrap analysis, for which we resampled the astrometric and RV data to create 100 synthetic data sets. Initially, we found the reduced $\chi^2$ for the RV data to be 69, clearly revealing systematic errors not accounted for in the original RV error analysis. In response to this, we treated all data with uniform error bars scaled to achieve a reduced $\chi^2$ of unity. The corresponding errors become thus $\pm 8.6$ km~s$^{-1}$ (resp.\ $\pm 1.35$ km~s$^{-1}$ and $\pm 3$ km~s$^{-1}$) for the H$\alpha$ set (resp.\ the primary and secondary {\it HST} data set from \citealt{Gies1998}). As mentioned earlier, we also chose to remove the epoch of 2012 Nov 06 from our astrometric fit because it has a much larger discrepancy with the orbital prediction than any other observation, and it must be an outlier due to unmodeled disk substructure or a calibration problem (we have included it in Fig.~\ref{fig:diskMIRC} for reference).

The result of our best-fit model along with 100 bootstraps is shown in Fig.~\ref{fig:diskMIRC}, the RV curve is shown in Fig.~\ref{fig:vr}.  The results on the orbital parameters, masses, and distance appear in Table~\ref{tab:orbit} and are discussed in Sect.~\ref{discussion}.

\begin{table}[t]
\label{tab:orbitalparameter}
\centering
\caption{Orbital elements of the \object{$\varphi$ Persei} binary system determined using interferometric and radial velocity measurements together, assuming a circular orbit. Note our orbital convention for the ephemeris here is referencing the location of the secondary relative to the primary (thus the primary star attains its most negative radial velocity at epoch $T_{\rm RV~min}$).} \label{tab:orbit}
\begin{tabular}{lr}
\hline
Parameter & Value \\ \hline\hline
 $T_{\rm RV~min}$ (RJD) & 56110.03$\pm$0.08 \\
 $P$ (d) & 126.6982 (fixed) \\
 $a$ (mas) & 5.89$\pm$0.02 \\
 $e$ & 0 (fixed) \\
 $i$ ($^\circ$) & 77.6$\pm$0.3 \\
 $\omega$($^\circ$) & 0 (fixed) \\
 $\Omega$($^\circ$) & -64.3$\pm$0.3 \\
 $\gamma$ (km\,s$^{-1}$) & -2.2$\pm$0.5 \\
 $K_a$ (km\,s$^{-1}$) & 10.2$\pm$1.0 \\
 $K_b$ (km\,s$^{-1}$) & 81.5$\pm$0.7 \\
    \hline
 $M_{a+b}$ ($M_\odot$) & 10.8$\pm$0.5 \\
 $M_{a}$   ($M_\odot$) & 9.6$\pm$0.3 \\
 $M_{b}$   ($M_\odot$) & 1.2$\pm$0.2 \\
    \hline
 $d (pc)$ & 186$\pm$3\\
\hline
\end{tabular}
\end{table}

\section{Visible data and disk properties}
\label{sec:Vega}
\subsection{Visible broadband data and model fitting}

Using the VEGA wide spectral band measurements, we first performed a model fitting using the software LITpro\footnote{LITpro software available at http://www.jmmc.fr/litpro}  \citep{litpro}. We first used the three bands containing continuum and then, separately, the band containing the H$\alpha$ line as defined in Table~\ref{tab:spectro}.\\

In a first approach, we just used a simple uniform disk model. We found a diameter in the continuum of $0.343\pm0.006$~mas and in the line of $0.423\pm0.012$ with a reduced $\chi^2$ of 9.3 and 14.4, respectively (Table~\ref{tab:litpro}, first section). The larger equivalent diameter in the line indicates the presence of an extended structure, corresponding to the line emission associated with the disk. The residuals clearly indicate that the size of the object varies with position angle, and thus we used a second model with the addition of an elongated Gaussian disk. The results are presented in Table~\ref{tab:litpro} (second section). We tried to detect the companion in the VEGA data but, despite the very good $(u,v)$ coverage, the data quality at low visibility level does not permit this direct detection. We therefore decided to adopt the position of the companion found by MIRC, but consider its flux as a free parameter. To avoid the complication introduced by the change in the companion's position, we only used the data of 2011 October 18 and 19 (RJD=55852 and 55853) for this analysis.

\begin{table}[ht]
\centering
\caption{\object{$\varphi$ Persei} model-fitting results for the continuum and the band containing H$\alpha$. The different models are 1) UD=equivalent uniform disk, 2) GD=elongated Gaussian disk, 3) UD+GD=central uniform disk of flux F1 + elongated Gaussian disk of flux F2, and 4) UD+GD+C=same as 3) + a companion of flux F3.}
\label{tab:litpro}
\setlength{\tabcolsep}{2pt}
\begin{tabular}{lcc}
& Continuum & H$\alpha$\\
\hline
\hline
UD & &\\
$\chi^2$& 9.3$\pm$0.1 & 14.4$\pm$0.1\\
$\theta$ & 0.343$\pm$0.006 & 0.423$\pm$0.012\\
\hline
GD & &\\
$\chi^2$ & 5.4$\pm$0.1 & 7.2$\pm$0.1 \\
$\theta_{min}$ & 0.141$\pm$0.006 & 0.182$\pm$0.010 \\
$elong$ & 2.34$\pm$0.14 & 2.32$\pm$0.18 \\
$PA$ & -78$\pm$1 & -72$\pm$2 \\
\hline
UD+GD & &\\
$\chi^2$ & 4.9$\pm$0.1&4.7$\pm$0.1\\
F1 & 0.47$\pm$0.13&0.45$\pm$0.08\\
$\theta$ & 0.1$\pm$0.1&0.1$\pm$0.1\\
F2 & 0.53$\pm$0.13&0.55$\pm$0.08\\
$\theta_{min}$ & 0.19$\pm$0.02&0.29$\pm$0.02\\
$elong$ & 2.64$\pm$0.49&2.89$\pm$0.29\\
$PA$ & -76$\pm$1&-69$\pm$2\\
\hline
UD+GD+C & &\\
$\chi^2$ & 4.3$\pm$0.1&4.2$\pm$0.1\\
F1 & 0.44$\pm$0.06&0.43$\pm$0.07\\
$\theta$ & 0.1$\pm$0.1&0.1$\pm$0.1\\
F2 & 0.53$\pm$0.06&0.54$\pm$0.09\\
$\theta_{min}$ & 0.18$\pm$0.01&0.28$\pm$0.02\\
$elong$ & 2.67$\pm$0.21&2.92$\pm$0.30\\
$PA$ & -76$\pm$1&-69$\pm$2\\
F3 & 0.033$\pm$0.006&0.029$\pm$0.008\\
\hline
\hline
\end{tabular}
\end{table}

The results clearly show the significant departure from a spherical object, both in the continuum and in the band containing H$\alpha$. As evidenced by the evolution of the reduced $\chi^2$, an improvement is found with a model consisting of a uniform disk + an elongated Gaussian disk (Table~\ref{tab:litpro}, third section), and the results are even better if we consider the companion (Table~\ref{tab:litpro}, fourth section). The difference in size and orientation of the elongated object between the continuum and H$\alpha$ is the signature of the difference between the disk emitting in the H$\alpha$ line and the continuum, and the elongated object
potentially consists of a rotationally distorted photosphere plus the contribution from the disk. In fact, a large portion of the stellar photosphere is obscured by the disk, but this fraction is not the same in the continuum and in the H$\alpha$ band. We also note that the position angle value found in the emission line region, where the disk is dominating the flux, is closer to the position angle found in the $H$ band ($PA=-64^\circ \pm 3^\circ$) and is compatible within 2~$\sigma$ despite the difference of wavelength.

\subsection{Spectrally resolved H$\alpha$ data and model fitting}
\label{sec:VegaDiff}
\begin{figure}[htbp]
  \centering
  \includegraphics[width=8cm]{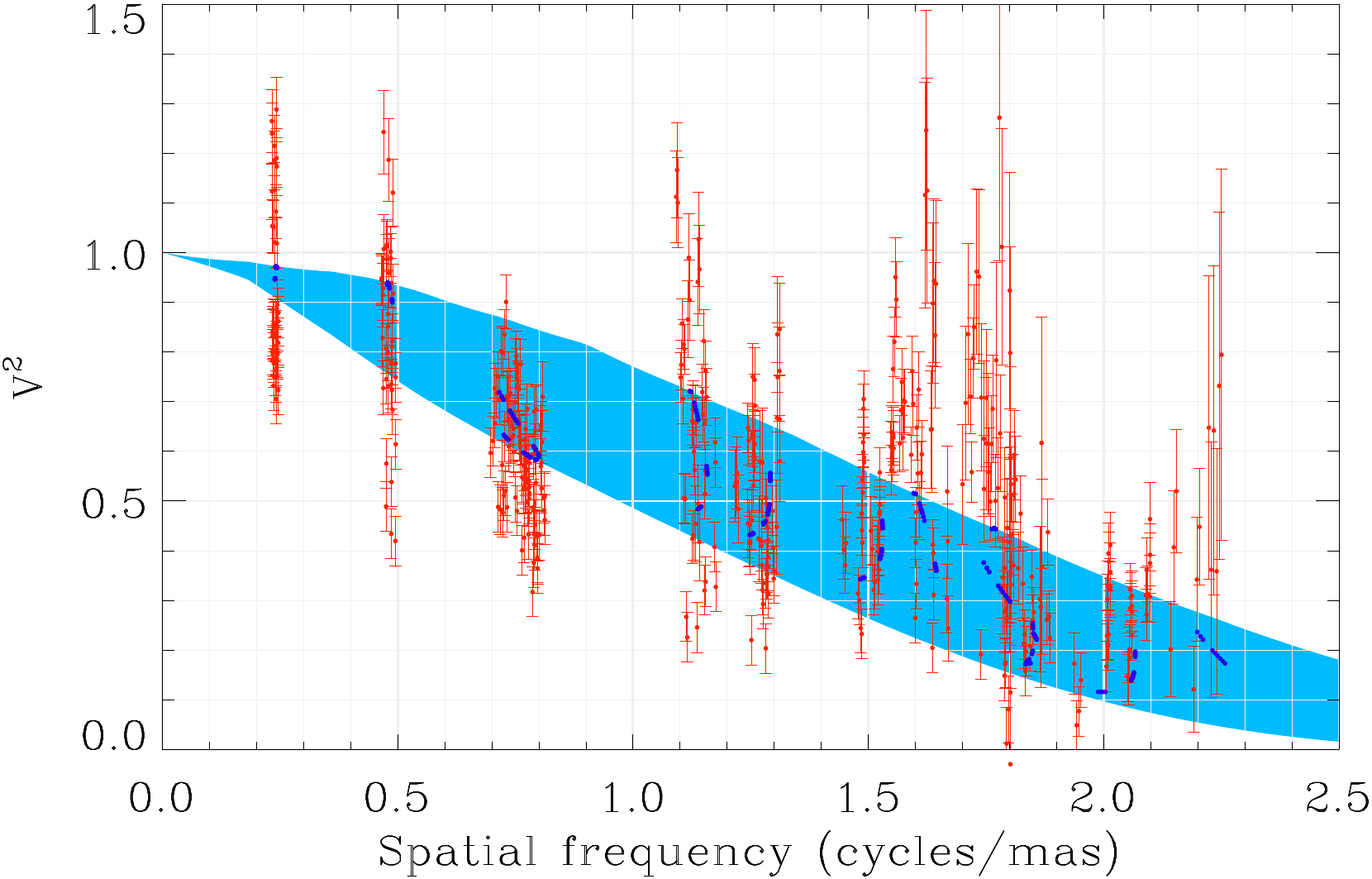}\\
  \includegraphics[width=8cm]{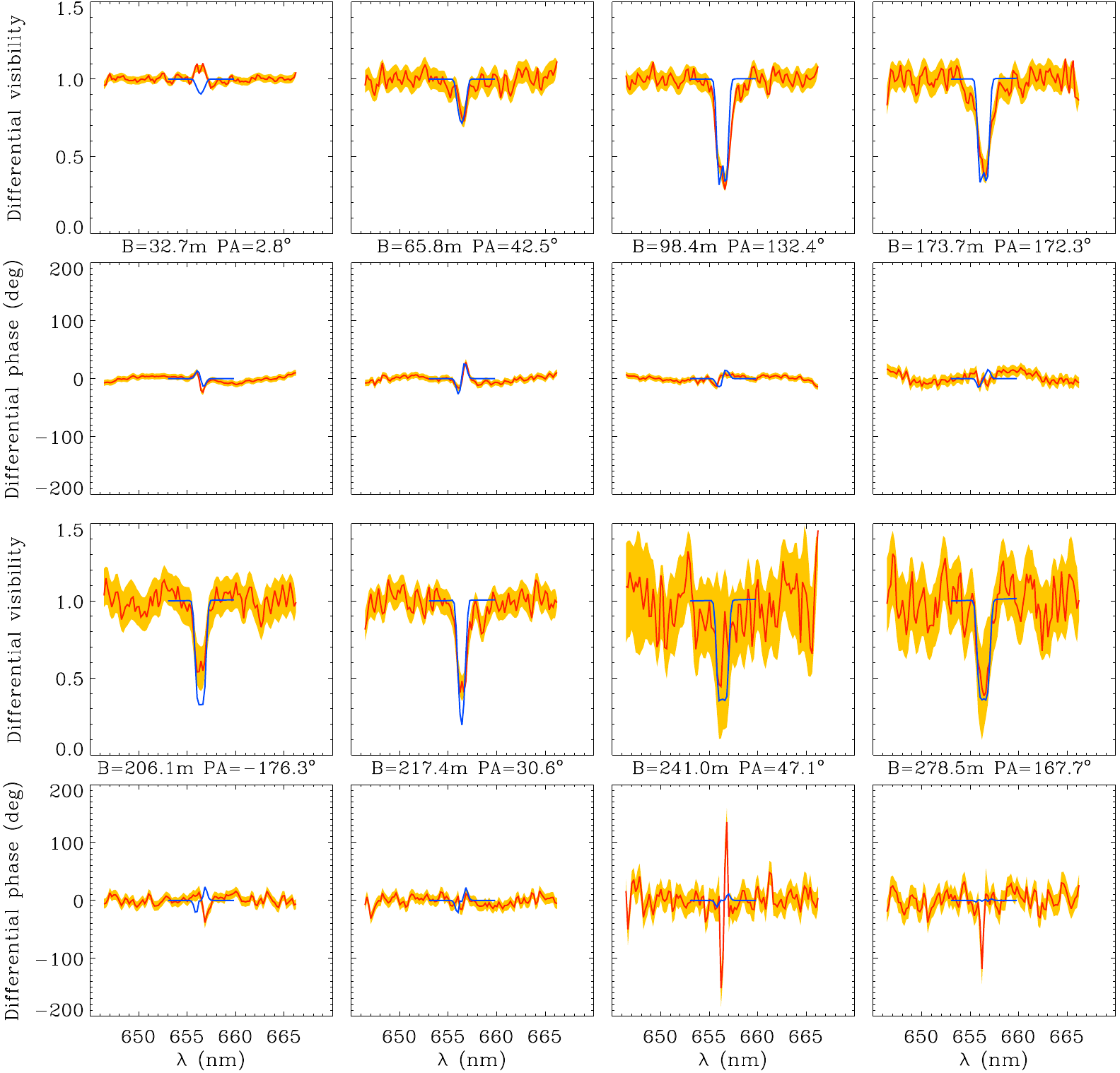}
  \caption{Top: 0.65$\mu$m continuum visibility for the best-fit model is plotted as a function of the spatial frequency, whereas the blue zone represents the extent of the model visibility for the various orientations of the baseline. Bottom: examples of differential visibilities and phases (in red + error bar in yellow) for the best-fit model (in blue) for various baselines between 32 and 278m.}
  \label{fig:VisVEGA}
\end{figure}

Differential visibilities are estimated through the standard VEGA procedure \citep{vega}. For the purpose of the \object{$\varphi$ Persei} work, we have established the absolute orientation of the differential phases in any of the CHARA+VEGA configurations \citep{vegaspie2012}. Some examples of differential measurements are presented in Fig.~\ref{fig:VisVEGA}, bottom part. The photon-counting detectors of VEGA suffer from a local saturation effect in presence of bright spectral lines that attenuates the line intensity. This nonlinearity in the photometric response of the detector does not generate bias in the derived complex visibilities as explained by \citet{Delaa2011}.

To model the spectrally resolved visibility and differential phase in the H$\alpha$ emission line, we used the simple kinematic model developed for fast model-fitting of a geometrically and optically thin equatorial disk in rotation described in detail by \citet{Delaa2011} and \citet{Meilland2012}.

The model parameters can be classified into four categories:
\begin{enumerate}
\item The stellar disk parameters: stellar polar radius ($R_\star$), stellar flattening (f), distance ($d$), inclination angle ($i$), and major-axis position angle ($PA$).
\item The kinematic parameters: rotational velocity ($V_{\rm rot}$) at the disk inner radius (i.e., photosphere) and the rotation power-law index ($\beta$).
\item The disk continuum parameters: disk FWHM in the continuum ($a_{\rm c}$) and disk continuum flux normalized by the total continuum flux ($F_{\rm c}$).
\item The disk emission line parameters: disk FWHM in the line ($a_{\rm l}$) and line equivalent width (EW).
\end{enumerate}

\begin{table}[!tbh]
\caption{Values of the VEGA best-fit kinematic model parameters.
\label{model_params}}
\centering
\begin{tabular}{ccc}
\hline
~~Parameter~~                    & ~~Value~~  &~~Remarks~~\\
\hline\hline
\multicolumn{3}{c}{\textbf{Global geometric parameters}}\\
$R_\star$                &    7$\pm$2 R$_\odot$                 & \\
$f$                            &    1.4$\pm$0.3                                  & \\
$d$                            &    186    pc         & this work\\
$i$                          & 78$^\circ$$\pm$10$^\circ$                             \\
$PA$                        & -68$^\circ$$\pm$ 5$^\circ$                 \\
\hline
\multicolumn{3}{c}{\textbf{Global kinematic parameters}}\\
$V_\mathrm{rot}$         &    450    $\pm$50 \,km\,s$^{-1}$            &\\
$\beta$                            &    0.5    $\pm$0.1                              & Keplerian rotation\\
\hline
\multicolumn{3}{c}{\textbf{Continuum disk geometry }}\\
$a_\mathrm{c}$     &    3.4$\pm$1.7        D$_\star$     & = 1.2$\pm$0.6 mas\\
$F_\mathrm{c}$    &    0.14$\pm$0.2     &\\
\hline
\multicolumn{3}{c}{\textbf{H$\alpha$ disk geometry }}\\
$a_\mathrm{H\alpha}$     &    5.9$\pm$1.7        D$_\star$             &= 2.1$\pm$0.6 mas\\
$EW_\mathrm{H\alpha}$    &    19.0$\pm$4.0            $\AA$                      &\\
\hline\hline
\end{tabular}

\end{table}

The model produces a data cube consisting of narrow spectral band images through the H$\alpha$ line at the instrumental spectral resolution. Visibilities and differential phases for our baseline configurations are then extracted from the data cube and are
compared to the observation using the standard $\chi^2$ method.

Except for the distance, the ten parameters are free and can vary within a range of realistic values. The parameter values for the best-fit model are presented in Table~\ref{model_params}. The reduced $\chi^2$ for this model is 4.0. Interestingly, with this more representative model of the system geometry, the position angle and inclination of the disk is fully compatible with the one found with the MIRC imaging.

The 0.65$\mu$m continuum-squared visibilities for the best-fit model are plotted as a function of the spatial frequency in Fig~\ref{fig:VisVEGA}, top part. Some examples of differential visibilities and phases for the best-fit model for various baselines between 32 and 278m are plotted in Fig~\ref{fig:VisVEGA}, bottom part. As already explained in \citet{Delaa2011}, phase variations for baselines overresolving the object (i.e., 60m in the major-axis orientation and 200m in the minor-axis one in the case of \object{$\varphi$ Persei}) are only poorly fitted. The jumps seen for some baselines might be either due to inhomogeneity in the disk or to VEGA data reduction biases at very low visibility. We also note that some of the very short baselines show an increase of visibility instead of a drop in the H$\alpha$ line, which corresponds to a smaller object in the line. This effect might be due to the high opacity of the disk seen edge-on and can only be modeled using a radiative transfer code.

\subsection{Spectrally resolved image reconstruction around H$\alpha$}
\label{sec:vegaImage}

\begin{figure*}[htbp]
  \centering
  \includegraphics[height=18cm,angle=-90,origin=br]{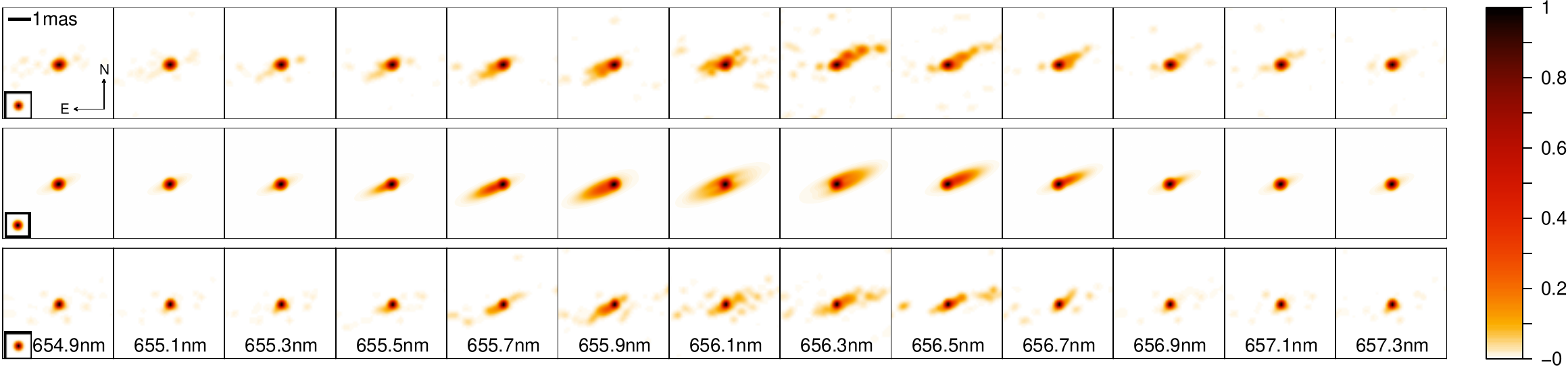}\\
  \vspace{6 mm}
\begin{tabular}{ccc}
  \includegraphics[width=5cm]{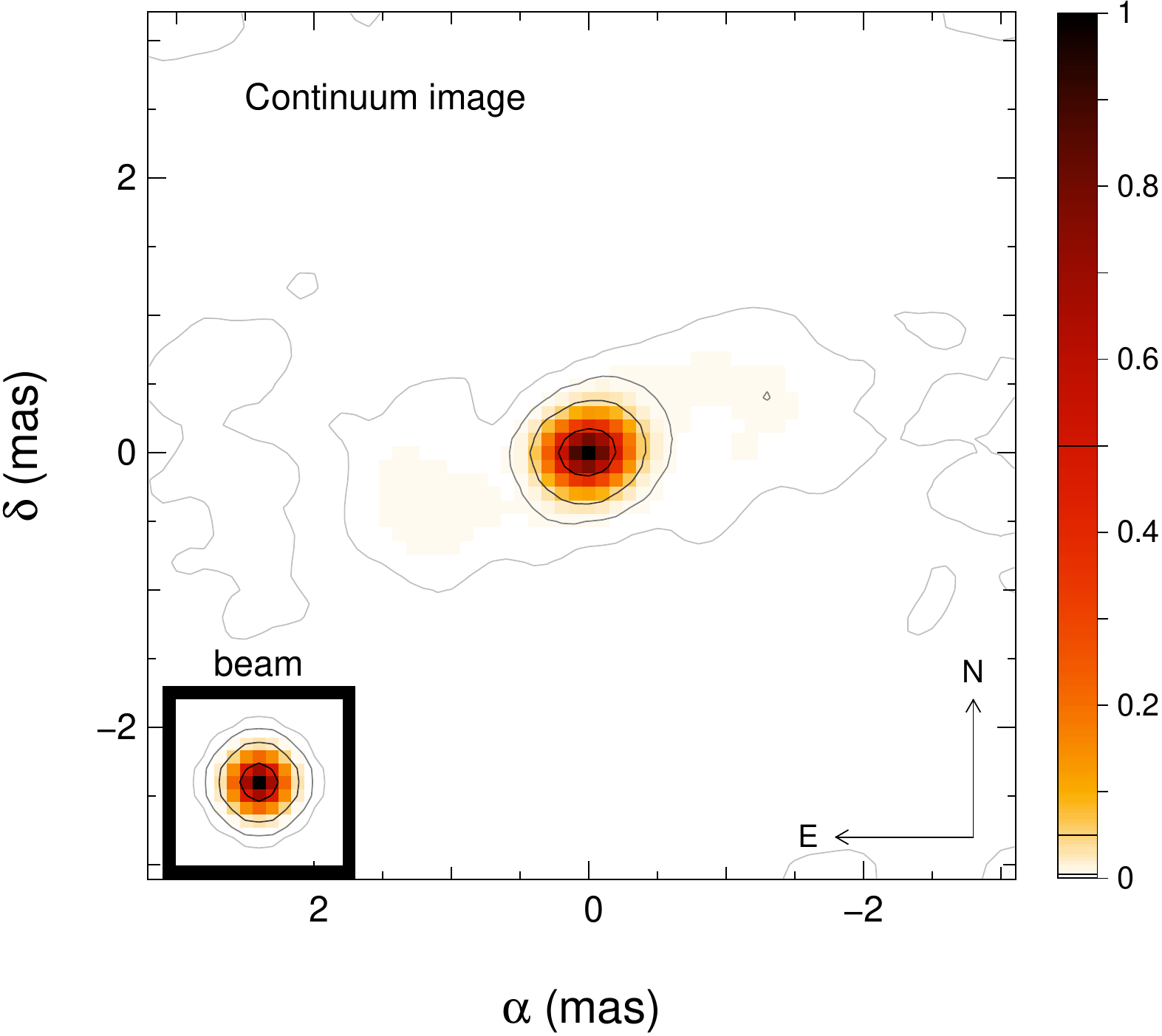}&
  \includegraphics[width=5cm]{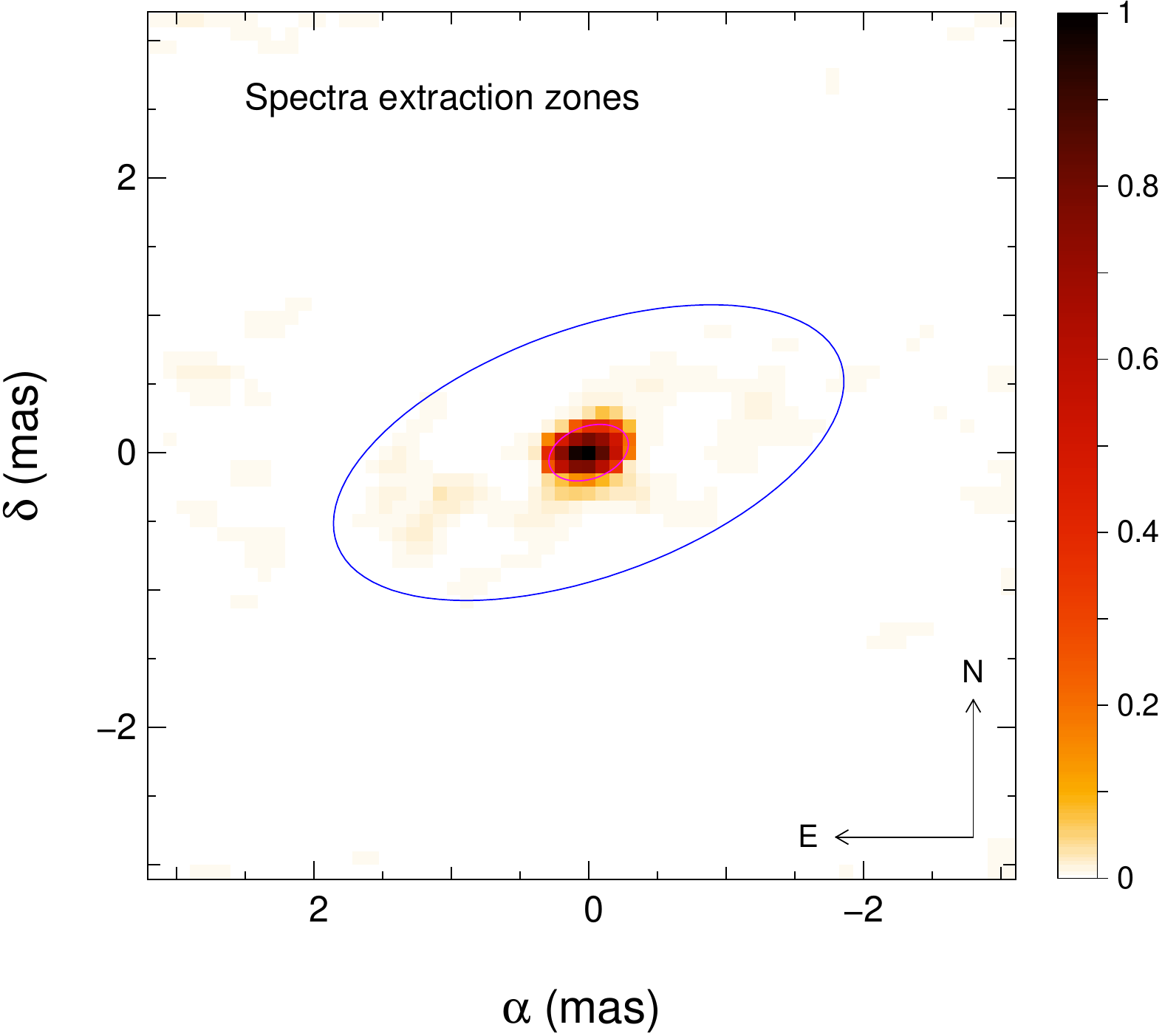}&
  \includegraphics[width=5cm]{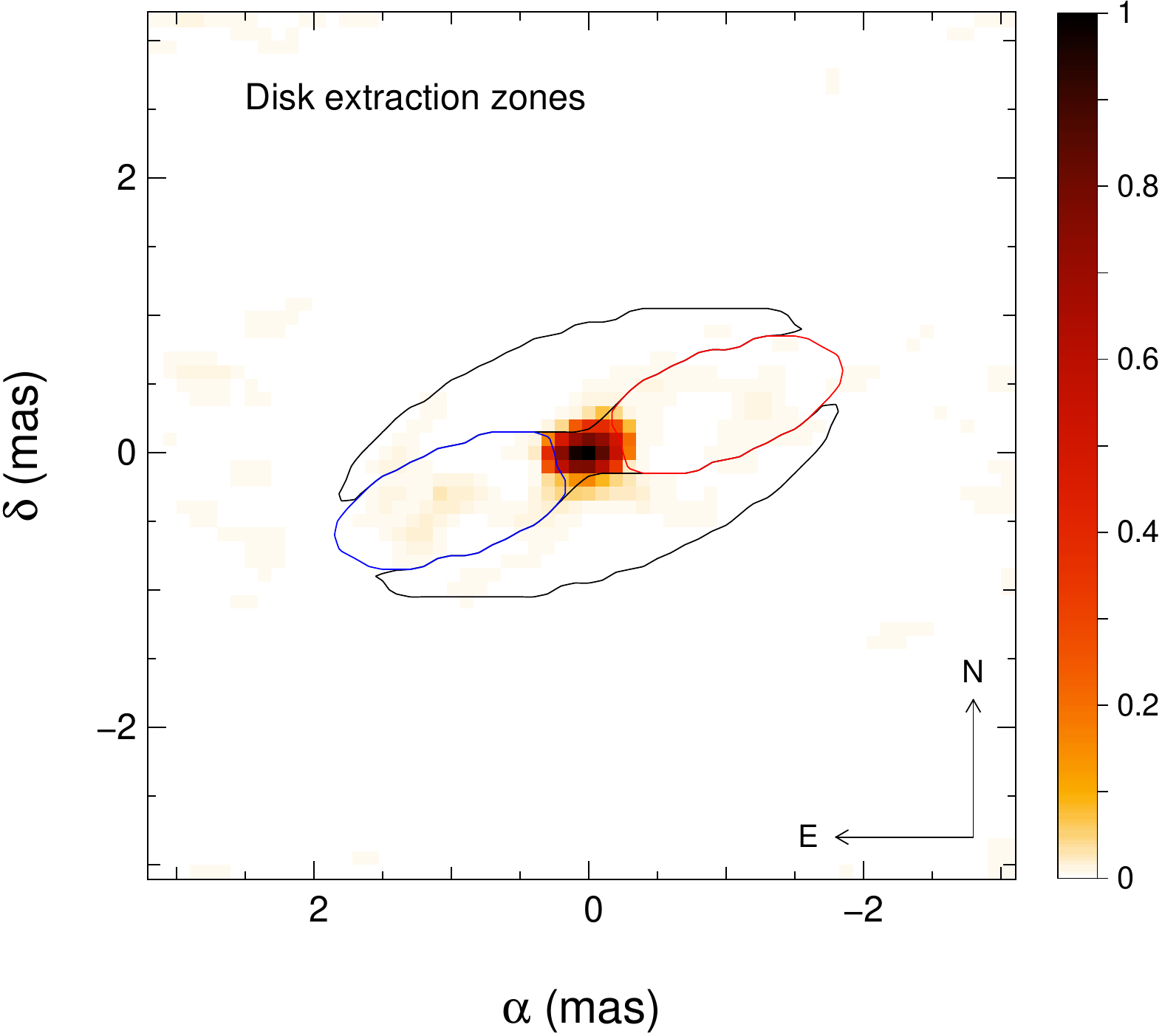}\\
 & \includegraphics[height=6cm]{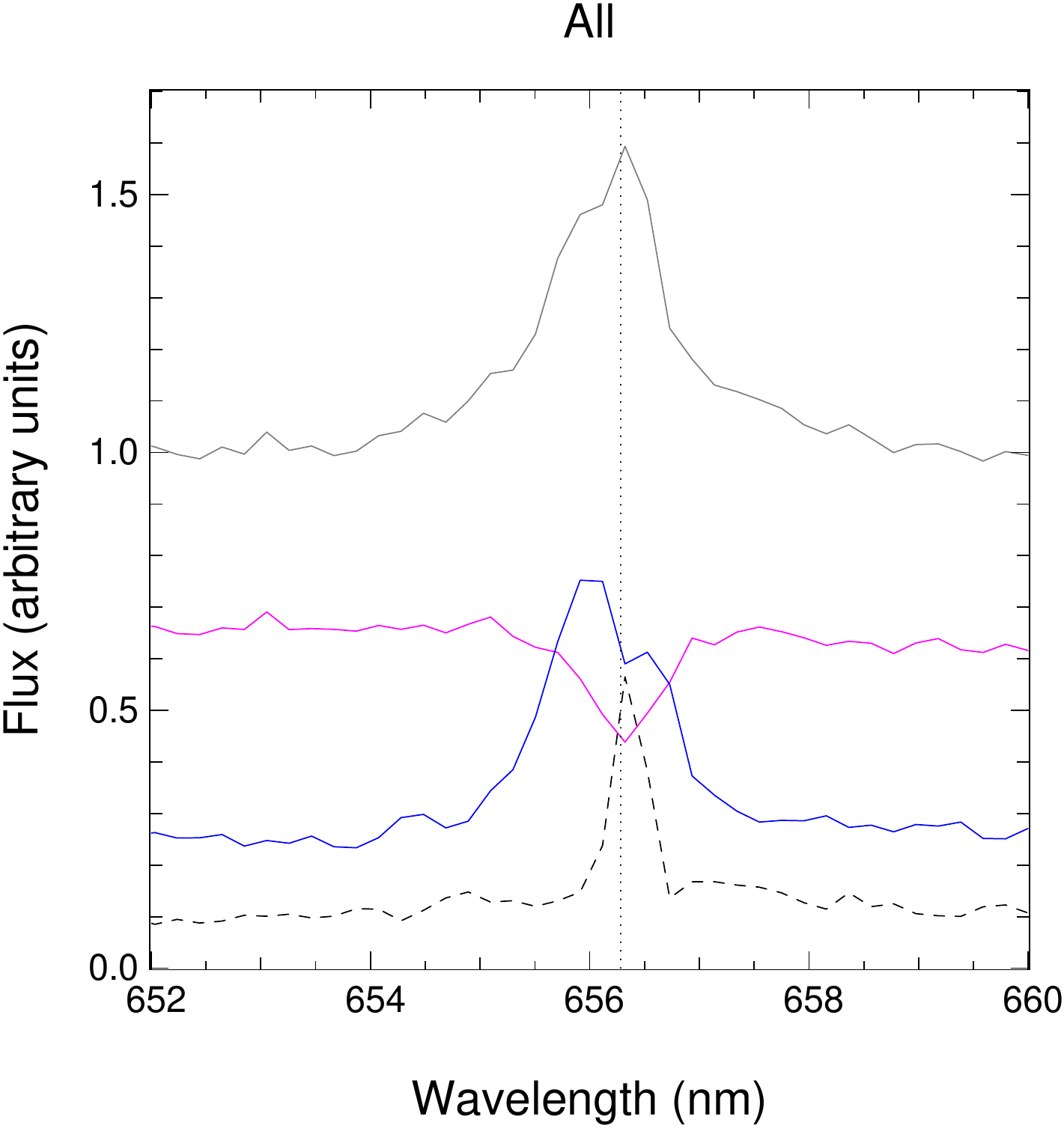}&
 \includegraphics[height=6cm]{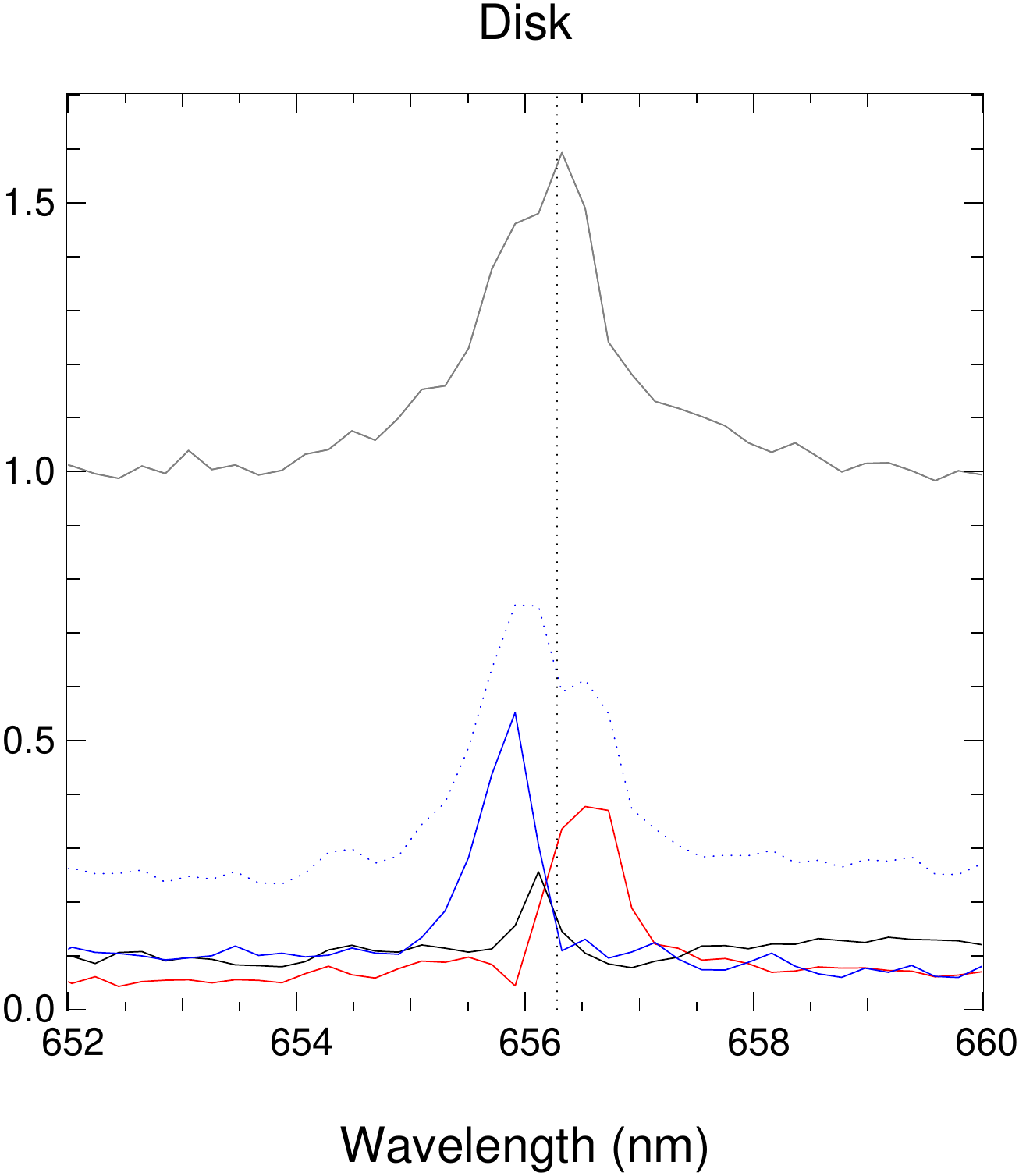}\\
\end{tabular}
  \caption{Top row: sequence of intensity maps (6.4x6.4mas) as a function of wavelength with north up and east left as imaged by VEGA (upper sequence), as computed from the kinematic model in Sect.~\ref{sec:VegaDiff} (middle sequence), and as imaged through a simulated observation of the above model (lower sequence). Middle row: at the left, we present the continuum image convolved to the diffraction limit of VEGA (0.2\,mas) together with contours at 50, 5, 0.5 and 0.05\% of the peak intensity to show the disk contour. The middle and right panels show the selected zones to extract the spectra presented below together with an unconvolved continuum image of the system.
 Bottom row: extracted spectra for the different spatial components: 1) Bottom middle: extracted fluxes (top: total flux, pink absorption feature: ``star'' zone, blue emission feature: ``disk'' zone, and dashed spectrum: all other contributions), 2) bottom right: decomposition of the ``disk'' zone: the top spectrum is the total flux. The blue and red lines correspond to the blue- and redshifted emission features, while the central zeroshifted emission feature comes from all other parts of the ``disk'' zone.}
  \label{fig:VegaImages}
\end{figure*}

We used the MiRA software \citep{Thiebaut2010} combined with an improved self-calibration method \citep{Millour2011} to reconstruct wavelength-dependent images of \object{$\varphi$ Persei} in the visible wavelength range. The data set contains the broadband $V^2$ defined in Table~\ref{tab:spectro} and the wavelength-dependent differential visibilities and phases, but no closure phases, as previously said. Therefore, the self-calibration method had to be modified. The different steps of our method, and the additions made in this work, are presented in Appendix~\ref{sec:reconstruction}. The so-called \emph{SELF-CAL} software and its documentation are available online\footnote{Available at http://self-cal.oca.eu}.

The image reconstruction leads to an $(x,y,\lambda)$ image cube
that covers a field of view of $6.4$\,mas with 64x64 pixels, a
wavelength range between 646\,nm and 666\,nm with a spectral
resolution of $\approx1640$. This translates into $\approx10$ spectral
channels in the H$\alpha$ emission line.

The images in Fig.~\ref{fig:VegaImages} exhibit little variation of morphology throughout the continuum. They display three prominent features:
\begin{itemize}
\item an elongated central source (the ``star'') that can be
  contained in an ellipse with $\approx0.6 \times 0.4$\,mas major and minor axes, and $\approx20^\circ$ position angle for the minor axis,
\item an extended flattened source (the ``disk'') whose extension is
  weakly constrained.  We can set an upper limit to its size to an ellipse $\approx3.9 \times 1.8$\,mas, and $\approx20^\circ$ position angle of the minor axis,
\item a significant amount of unconstrained flux
  ($\approx15\%$ of the continuum flux, throughout all wavelengths),
  which could be linked to the companion star or to
  image-reconstruction artifacts.
\end{itemize}

The images in the emission line are very different from those in the continuum, showing emission that sweeps across the ``disk'' feature. In the upper part of Fig.~\ref{fig:VegaImages}, we compare the spectral images and the maps deduced from the model presented in the previous section. We also provide a set of reconstructed images from simulated data based on the kinematic model.

To extract information from these images from milli-arcsecond
integral-field spectroscopy, we multiplied the
images by the system spectrum (as MiRA provides total-flux-normalized
images) and selected different zones with binary masks to
extract their spectra. The result of this process is shown in the bottom part of Fig.~\ref{fig:VegaImages}.

The ``star'' zone exhibits an absorption spectrum, characteristic of a
stellar spectrum. The ``disk'' zone bears most of the emission of the
system, while the unconstrained flux has a flat spectrum.

The results of the VEGA imaging and model fitting are discussed in general in Sect.~\ref{discussion}.

\section{Discussion}
\label{discussion}

\subsection{Fundamental stellar parameters}
We first consider the dimensions of the binary system.
The astrometric measurements from MIRC (Table~\ref{tab:companionMIRC})
were combined with the radial velocity measurements for the two
components (Table~\ref{tab:rv}) in a joint solution for the binary
orbital elements and distance that is summarized in Table~7. Our derived values of the stellar masses,
$M_a = 9.6 \pm 0.3 M_\odot$ and $M_b = 1.2 \pm 0.2 M_\odot$, agree well with those determined previously by \citet{Gies1998}.
Our distance estimate $d=186 \pm 3$~pc is about $4.2\sigma$ closer than
the estimate from the revised {\it Hipparcos} distance
or $1.1\sigma$ closer than the original {\it Hipparcos} distance.
We note, however, that \object{$\varphi$ Persei} is flagged as ''unsolved variable''
in the {\it Hipparcos} catalog; we could then expect an underestimated
error for the parallax in the revised {\it Hipparcos} catalog.

A knowledge of the distance to \object{$\varphi$ Persei} has consequences
for the size of the Be star. We can estimate the
stellar radius from the predicted angular size and distance.
We first consider the angular size from a comparison from a
of the observed and predicted photospheric fluxes,
a task that is best accomplished in the ultraviolet part of the spectrum
where the circumstellar disk flux is negligible. \citet{Touhami2013} fit the observed UV spectrum of \object{$\varphi$ Persei} using a Kurucz flux model for the adopted temperature (from \citet{Fremat2005}) of the Be star and a standard extinction law (ignoring the small flux contribution of the hot companion), and
they derived a limb-darkened, angular diameter for the Be star
of $0.235 \pm 0.008$ mas. This is similar to the minor-axis diameter
derived from the VEGA broadband observations ($0.18 \pm 0.01$ mas, Table~\ref{tab:litpro}). Then, for an adopted distance of 186 pc, the stellar radius is estimated to be $R_a / R_\odot = 4.7 \pm 0.4$.  This is somewhat lower but consistent within error bars to that given in Table~\ref{model_params} and prorated for the adopted lower distance ($R_a / R_\odot = 5.9 \pm 1.7$). The difference may be partially due to the neglect of UV flux scattered away from the line of sight by electron scattering in the inner disk so that the stellar UV flux (and hence angular size) is somewhat higher than estimated from the observed UV spectrum.  Furthermore, the estimate of stellar diameter from the VEGA measurements may be influenced by the assumed flux distribution in the disk.

The mass and radius of the Be star match the mass -- radius relationship
for main-sequence stars derived from eclipsing binaries \citep{Torres2010}.  Furthermore, the effective temperature $T_{\rm eff} = 29.3$~kK \citep{Gies1998} and luminosity $\log L_a/L_\odot = 4.16 \pm 0.10$ (for $R_a / R_\odot = 4.7$) are consistent with those of eclipsing binary stars of similar mass \citep{Torres2010}. The Be star's rotation rate, however, is exceptional.  Adopting the estimate of projected rotational velocity from \citet{Fremat2005}, $V\sin i = 462 \pm 33$ km~s$^{-1}$, we derive a ratio of
equatorial velocity to orbital velocity at the equator \citep{Rivinius2013} of
$(V\sin i / \sin i) / V_{\rm crit} = 0.93 \pm 0.08$ for the
mass and radius given above and an inclination from Table~\ref{tab:MIRCmodel}
(assuming coalignment of the spin and orbital vectors; see below).
Note that we have tacitly assumed that the radius derived from the
UV flux and the distance corresponds to the star's polar radius,
whereas it may represent some value between the polar and (larger)
equatorial radius.  For example, if our radius estimate corresponds to
the equatorial radius in a near critical rotation star, then
$R_{\pm polar} = R_a / 1.5$ and $(V\sin i / \sin i) / V_{\rm
crit} = 0.76 \pm 0.08$.
On the other hand, the measured projected velocity may underestimate
the actual value of $V\sin i$ because of gravity darkening
\citep{Townsend2004}. Despite these uncertainties, the clear conclusion is that the Be star is rotating very rapidly. The companion's rotational velocity is low, $v_{\rm rot} \sin i < 10$ km~s$^{-1}$ \citep{Gies1998}.

\citet{Gies1998} derived a radius ratio of $R_b / R_a = 0.20$ from the
observed flux ratio in the ultraviolet spectrum.  Adopting the Be star
radius given above, the radius and luminosity of the hot companion
are $R_b / R_\odot = 0.94 \pm 0.09$ and $\log L_b /L_\odot = 3.80 \pm 0.13$
(for $T_{\rm eff} = 53 \pm 3$~kK; \citet{Gies1998}).  Thus, although the
Be star is five times larger than the sdO companion, it is only twice as
luminous.  In the Rayleigh-Jeans tail of the spectrum in the near-IR, we would expect that the surface flux varies with effective temperature, so the flux ratio would be $F_b / F_a \approx (T_b / Ta)(R_b / R_a)^2$. Based upon the temperature and radius ratios estimated by \citet{Gies1998}, the predicted optical/IR flux ratio is $8\%$. This value is similar to the observed flux ratio in the optical ($F3/F1 = 7.5 \pm 1.7 \%$; Table~\ref{tab:litpro}), but higher than found in the $H$ band ($f_2/f_1 = 2.1 \pm 0.5\%$; Table~\ref{tab:MIRCmodel}). We speculate that the discrepancy is due to obscuration of the sdO companion by circumstellar gas that becomes more significant at longer wavelength (see below).

\subsection{Birthplace and age}
We next consider the possible birthplace and age of \object{$\varphi$ Persei}.
Our derived distance is consistent with that of the nearby \object{$\alpha$~Per} cluster (Melotte 20; Per OB3) at 177 pc \citep{Zeeuw1999}, and
\object{$\varphi$ Persei} is located at a similar angular separation ($15.9^\circ$)
from the cluster center as some of the $\alpha$~Per ``halo stars''
identified by \citet{Zeeuw1999}. Furthermore, the proper motion of \object{$\varphi$ Persei} ($\mu_\alpha \cos \delta = 24.6$ mas~yr$^{-1}$,
 $\mu_\delta = -14.0$ mas~yr$^{-1}$; \citealt{leeuw2007b}) is
similar to that of the $\alpha$~Per cluster ($\mu_\alpha \cos \delta = 22.7$ mas~yr$^{-1}$,  $\mu_\delta = -26.5$ mas~yr$^{-1}$; \citealt{Lodieu2012}).
Likewise, the systemic radial velocity of \object{$\varphi$ Persei} ($-2.2$ km~s$^{-1}$; Table~7) is similar to the mean value for the $\alpha$~Per cluster ($-1.0$ km~s$^{-1}$; \citealt{Zeeuw1999}). Thus, the distance, position, and kinematical properties of \object{$\varphi$ Persei} suggest that it may be an outlying member of the $\alpha$~Per cluster. If so, then the age of \object{$\varphi$ Persei} is the cluster age, $\approx 52$~Myr \citep{Kharchenko2005,Makarov2006, Lyubimkov2010}. This is much older than the main-sequence lifetime of $\approx 25$~Myr for a $10 M_\odot$ mass star like the Be primary \citep{Brott2011},
which suggests that the Be star attained its high mass relatively recently.

The age of the $\alpha$~Per cluster is consistent with the binary
evolutionary scenario for \object{$\varphi$ Persei}.  The \object{$\varphi$ Persei} binary probably
began its life as a pair of $6 M_\odot + 5 M_\odot$ stars with a shorter
orbital period in the range of 7~d \citep{Vanbeveren1998b} to
14~d \citep{Vanbeveren1998}.  As the initially more massive star
completed core-hydrogen burning, it expanded and began mass and
angular momentum transfer to the companion.  The binary separation
increased following mass ratio reversal, reaching its current state
of wide separation between the stripped-down remnant of the mass donor
(the hot subdwarf star) and the rapidly rotating mass gainer (the Be star).
If we assume that the mass transfer was approximately conservative
(i.e., negligible mass lost from the system), then the progenitor
of the hot subdwarf had a mass at least as high as half the current
total mass of the system ($5.4~M_\odot$; Tab.~7).
Indeed, given the current advanced evolutionary state of the donor as a hot subdwarf star, it must have had an original mass in excess of that for any H-burning star still in the $\alpha$~Per cluster.  The most massive star
in the $\alpha$~Per cluster (aside from \object{$\varphi$ Persei}) is $\alpha$~Per itself, with a mass estimated in the range of $6.7 M_\odot$ \citep{Makarov2006} to $7.3 M_\odot$ \citep{Lyubimkov2010}. Thus, the original mass of the donor star was probably $M_{b~{\rm orig}}\approx 7 M_\odot$, so that the mass gainer (Be star) accreted most of the difference $M_{b~{\rm orig}} - M_b \approx 6 M_\odot$
during mass transfer.  This wholesale transformation of the mass gainer
led to its current high mass and fast spin, and the Be disk
found today represents the Be star's means to shed its excess
angular momentum. We now know of two other Be plus sdO binaries,
\object{FY~CMa} and \object{59~Cygni}, that share many of the properties of \object{$\varphi$ Persei}
\citep{Peters2013}, and it is probable that other hot companions
of Be stars await discovery \citep{Koubsky2012,Koubsky2014}.

\subsection{Properties of the Be-star disk}
We next consider the properties of the Be star disk.
Interferometric observations have now resolved the disk in
the continuum at several wavelengths.  The angular size
derived depends in part on the adopted model for the
flux distribution in the sky, but we list in Table~\ref{tab:interfero}
the results for an assumed uniform disk star and
Gaussian elliptical disk.  The Gaussian FWHM along the
disk major axis $\theta_{\rm maj}$ appears to increase with
wavelength from $R$ through $H$ and $K'$ bands as expected
because of the increase in hydrogen free-free opacity with wavelength
that causes the optical depth unity radius to appear larger
when moving into the near-infrared \citep{Touhami2011}.
The H$\alpha$ diameter measured from the VEGA observations
is somewhat smaller than that derived from earlier H$\alpha$ measurements
and is comparable to or lower than the contemporaneous near-IR continuum estimates.
This is surprising given that the H$\alpha$ sizes are generally larger
than the near-IR continuum sizes in most Be stars \citep{Touhami2013}.  We note, however, that the H$\alpha$ emission equivalent
width at the time of the VEGA observations ($W_\lambda=-25$\AA)
was much weaker than at some times in the past ($W_\lambda=-46$\AA~ in 2001), and if the disk size scales with the emission strength
\citep{Grundstrom2006}, then the disk dimensions may have
been larger at earlier epochs.  It is possible that gravitational
interactions with the companion limit the physical boundary of the
gaseous disk, as occurs in many Be X-ray binaries
\citep{Negueruela2001}.

The interferometric observations indicate that the spin angular momentum axis of the disk is coaligned with the orbital angular momentum axis.
The position angle of the long axis of the projected disk ($PA$ in Tables~\ref{tab:interfero}, \ref{tab:MIRCmodel}, \ref{tab:litpro}, and \ref{model_params}) agrees with the longitude of the ascending node in the astrometric orbital solution ($\Omega$ in Table~7). The position angle of the disk normal ($PA + 90^\circ$) agrees with the intrinsic polarization angle, as expected by \citet{Quirrenbach1997}. Furthermore, the disk inclination derived from the ratio of projected minor to major axis ($i= 75^\circ \pm 9^\circ$ from $r= \cos i$ in Tables~\ref{tab:interfero} and \ref{model_params}) is consistent within errors with the astrometric orbital inclination ($i=77\fdg6 \pm 0\fdg3$ in Table~7). Finally, the sense of the disk rotation (moving away in the northwestern part; Fig.~\ref{fig:VegaImages}) is the same as the sense of orbital motion (Fig.~\ref{fig:diskMIRC}). Thus, we find that
the disk and orbital angular momentum vectors are coaligned, verifying the
prediction from the binary interaction model that the spin-up of the Be star
was caused by mass and angular momentum transfer from the donor companion.
Note that in general we expect that the inclination derived from
the projected shape of the disk will be smaller than actual because of disk flaring that increases the apparent size of the minor axis \citep{Grundstrom2006}, so the small difference we find between inclination estimates from the disk and orbit is probably consistent with some disk flaring. The increase in disk thickness with radius (flaring) means that there will be a range in inclination below $90^\circ$ where the disk may occult our view of the Be star photosphere, and this may explain the presence of narrow ``shell'' lines that are observed in the spectra of some Be stars \citep{Rivinius2013}. \object{$\varphi$ Persei} has a well-observed shell line spectrum, so our inclination estimate would appear to indicate that the disk half-opening angle may amount to as much as $90^\circ - i = 12^\circ$ in this case.  Curiously, \citet{Hanuschik1996} used a statistical study of the numbers of Be-shell and normal Be stars to arrive at a similar half-opening angle of $13^\circ$. The agreement may be coincidental, and detailed models of disk line formation show that shell line spectra may form over a wider range below $i=90^\circ$ \citep{Silaj2014}.

Next, we offer a few comments about the surface brightness distribution of
the Be disk. Similar to other investigators, we have used Gaussian elliptical models
for surface brightness because they generally provide good fits of the
interferometric visibility. However, models of the spatial appearance of disk flux often predict more complicated distributions (Fig.~\ref{fig:VegaImages}), and in particular, in models where the circumstellar gas becomes optically thick, the surface brightness distribution appears more or less constant in the parts of the disk close to the photosphere \citep{Gies2007}. Thus, for example, the relatively small minor axis fit to the MIRC observations (Table~\ref{tab:MIRCmodel}) may be partially due to the non-Gaussian appearance of the foreshortened disk brightness distribution (Fig.~\ref{fig:diskMIRC}). It is probably unrealistic to assume that the disk is azimuthally symmetric, because many Be stars including \object{$\varphi$ Persei} display long-lived asymmetries in their emission line profiles that probably form in one-armed spiral structures \citep{Rivinius2013}. The H$\alpha$ profile observed over the time span of the CHARA observations displayed a stronger red than blue peak (see Fig.~8 of \citet{Silaj2010}), which indicates that some asymmetry in the H$\alpha$ disk surface brightness did exist at this time.  We note that there is a hint in the H$\alpha$ image reconstructions shown in Fig.~\ref{fig:VegaImages} that the northwestern (redshifted) extension of the disk is slightly brighter than the southeastern extension (blueshifted). The VEGA observations were obtained near orbital phase 0.0 when the companion appears near its northwestern, maximum elongation, and the part of the disk facing the companion may be brighter \citep{Hummel2001}. Thus, the H$\alpha$ interferometry offers tentative evidence for the presence of the surface brightness asymmetries expected from disk density structures and heated zones.

Finally we note that reverse mass transfer from the Be disk to the hot
subdwarf may occur. The fact that the observed circumstellar disk radius appears smaller than the semimajor axis (Fig.~\ref{fig:diskMIRC}) does not necessarily imply that disk gas never reaches the vicinity of the companion. The interferometric observations record the spatial distribution of the brightest parts of the disk emission, and it is possible that lower density and fainter disk gas does extend to the orbit of the companion as it slowly moves outward from the Be star.  Indeed, \citet{Hummel2003} argued that the He~I $6678$\AA~ emission extends to the outer edge of the disk at $R_{\rm Roche} / a = 0.56$.  We would expect that some of this gas will be captured by the gravitational pull of the companion to form an accretion disk around the subdwarf star. The double-peaked appearance of the He~II $4686$\AA~ emission that forms near the companion may be evidence of such an accretion
disk \citep{Poeckert1981}. It is possible that gas around the
companion may obscure our view of the hot subdwarf star itself,
and the attenuation would increase with wavelength if free-free
processes dominate the continuum opacity. This may explain
why the companion appears fainter than expected in the near-IR.

\section{Conclusion}
\label{conclusion}
Thanks to advanced visible and infrared interferometric instrumentation at the CHARA Array, we have been able to obtain the first high angular resolution images of the Be binary \object{$\varphi$ Persei}, including the first spectrally resolved images in the visible (made with VEGA in four-telescope mode).  These images show the highly inclined circumstellar disk surrounding the Be star, and the helium star remnant of the mass donor star is detected for the first time through MIRC observations of its orbital motion.

The relative angular orbital motion of the hot companion was combined with new orbital radial velocity measurements of both components to derive a full, three-dimensional orbit of the binary system.  We obtain masses of $9.6 \pm 0.3 M_\odot$ and $1.2 \pm 0.2 M_\odot$ for the Be star primary and subdwarf secondary, respectively. The distance derived from the joint astrometric and spectroscopic solution is $186 \pm 3$~pc. We note that \object{$\varphi$ Persei} shares the same distance, kinematical, and evolutionary properties as members of the $\alpha$~Per star cluster, and if this association is confirmed, then the cluster age will place significant constraints on the initial masses and evolutionary mass transfer processes that transformed the \object{$\varphi$ Persei}
binary system. The Be star (mass gainer) has a mass, radius, and
effective temperature that are typical of an early B-type main-sequence
star, but it is rotating very rapidly ($93 \pm 8 \%$ of the critical rate) as
a consequence of angular momentum gained during past mass transfer. The sdO secondary star (remnant of the mass donor) is fainter in the near-IR than predicted from the FUV flux ratio \citep{Gies1998}, due perhaps to obscuration by circumstellar gas in its vicinity.

The circumstellar gas disk of the Be star primary was resolved in the optical
and near-IR $H$-band, and we estimated the disk flux contributions
in each of the H$\alpha$ emission lines and the optical and near-IR continua.  The apparent ellipticity and position angle of the disk in the sky is similar in each band.  The size of the disk in the continuum flux is larger at longer wavelength, presumably due to the increasing free-free optical depth of the disk at longer wavelength. The disk appears largest in the spectrally resolved H$\alpha$ observations (Fig.~\ref{fig:VegaImages}), but the visible flux appears to be confined well within the Roche lobe of the Be star. We find that the angular momentum vector of the disk is aligned with and shows the same sense of rotation as the binary orbital angular momentum vector.  This agreement confirms the evolutionary prediction that the Be star was spun up by mass transfer in the binary and that this excess angular momentum is now being shed into the Be star disk.

The appearance of the Be disk is presented in the first nonparametric images
reconstructed from the MIRC $H$-band observations (Fig.~\ref{fig:diskMIRC}) and from the spectrally dispersed VEGA observations of H$\alpha$ (Fig.~\ref{fig:VegaImages}). These images reveal the brighter, inner
parts of the disk, and there is some evidence in the spectrally
dispersed H$\alpha$ maps from VEGA of an asymmetry in the azimuthal
light distribution of the circumstellar disk. The northwestern
part of the disk that is moving away from us appears slightly
brighter than the southeastern part that is approaching (Fig.~\ref{fig:VegaImages}),
consistent with the relatively stronger red peak of the H$\alpha$
profile observed at this time. Such asymmetries probably originate
in long-lived disk structures (perhaps a one-armed spiral instability),
and long-baseline interferometric observations now appear to offer
the means of exploring the time evolution of these disk processes.

\begin{acknowledgements}
The CHARA Array is operated with support from the National Science Foundation through grant AST-0908253, the W. M. Keck Foundation, the NASA Exoplanet Science Institute, and from Georgia State University. VEGA is a collaboration between CHARA and OCA-Lagrange/UJF-IPAG/UCBL-CRAL that has been supported by the French programs PNPS and ASHRA, by INSU and by the R\'{e}gion PACA. We acknowledge the use of the electronic database from the CDS, Strasbourg and electronic bibliography maintained by the NASA/ADS system. Part of this material is based upon work supported by the National Science Foundation under Grant No.~AST-1009080 (DRG) and grants NSF-AST0707927 and NSF-AST1108963 (JDM). We finally warmly thank Dietrich Baade, our referee, for his accurate review.
\end{acknowledgements}

\bibliographystyle{aa}
\bibliography{phiperph}

\begin{thebibliography}{80}
\expandafter\ifx\csname natexlab\endcsname\relax\def\natexlab#1{#1}\fi

\bibitem[{{Abt} {et~al.}(2002){Abt}, {Levato}, \& {Grosso}}]{Abt2002}
{Abt}, H.~A., {Levato}, H., \& {Grosso}, M. 2002, \apj, 573, 359

\bibitem[{{Bonneau} {et~al.}(2006){Bonneau}, {Clausse}, {Delfosse}, {Mourard},
  {Cetre}, {Chelli}, {Cruzal{\`e}bes}, {Duvert}, \& {Zins}}]{SearchcalBright}
{Bonneau}, D., {Clausse}, J.-M., {Delfosse}, X., {et~al.} 2006, \aap, 456, 789

\bibitem[{{Bonneau} {et~al.}(2011){Bonneau}, {Delfosse}, {Mourard}, {Lafrasse},
  {Mella}, {Cetre}, {Clausse}, \& {Zins}}]{SearchcalFaint}
{Bonneau}, D., {Delfosse}, X., {Mourard}, D., {et~al.} 2011, \aap, 535, A53

\bibitem[{{Bo\v{z}i\'{c}} {et~al.}(1995){Bo\v{z}i\'{c}}, {Harmanec}, {Horn},
  {Koubsky}, {Scholz}, {McDavid}, {Hubert}, \& {Hubert}}]{Bozic1995}
{Bo\v{z}i\'{c}}, H., {Harmanec}, P., {Horn}, J., {et~al.} 1995, \aap, 304, 235

\bibitem[{{Brott} {et~al.}(2011){Brott}, {de Mink}, {Cantiello}, {Langer}, {de
  Koter}, {Evans}, {Hunter}, {Trundle}, \& {Vink}}]{Brott2011}
{Brott}, I., {de Mink}, S.~E., {Cantiello}, M., {et~al.} 2011, \aap, 530, A115

\bibitem[{{Campbell}(1895)}]{Campbell1895}
{Campbell}, W.~W. 1895, \apj, 2, 177

\bibitem[{{Che} {et~al.}(2012{\natexlab{a}}){Che}, {Monnier}, {Kraus}, {Baron},
  {Pedretti}, {Thureau}, \& {Webster}}]{mirc2012}
{Che}, X., {Monnier}, J.~D., {Kraus}, S., {et~al.} 2012{\natexlab{a}}, in SPIE
  Conference Series, Vol. 8445

\bibitem[{{Che} {et~al.}(2012{\natexlab{b}}){Che}, {Monnier}, {Tycner},
  {Kraus}, {Zavala}, {Baron}, {Pedretti}, {ten Brummelaar}, {McAlister},
  {Ridgway}, {Sturmann}, {Sturmann}, \& {Turner}}]{che2012}
{Che}, X., {Monnier}, J.~D., {Tycner}, C., {et~al.} 2012{\natexlab{b}}, \apj,
  757, 29

\bibitem[{{Clarke} \& {Bjorkman}(1998)}]{Clarke1998}
{Clarke}, D. \& {Bjorkman}, K.~S. 1998, \aap, 331, 1059

\bibitem[{{de Mink} {et~al.}(2013){de Mink}, {Langer}, {Izzard}, {Sana}, \& {de
  Koter}}]{deMink2013}
{de Mink}, S.~E., {Langer}, N., {Izzard}, R.~G., {Sana}, H., \& {de Koter}, A.
  2013, \apj, 764, 166

\bibitem[{{de Zeeuw} {et~al.}(1999){de Zeeuw}, {Hoogerwerf}, {de Bruijne},
  {Brown}, \& {Blaauw}}]{Zeeuw1999}
{de Zeeuw}, P.~T., {Hoogerwerf}, R., {de Bruijne}, J.~H.~J., {Brown}, A.~G.~A.,
  \& {Blaauw}, A. 1999, \aj, 117, 354

\bibitem[{{Delaa} {et~al.}(2011){Delaa}, {Stee}, {Meilland}, {Zorec},
  {Mourard}, {B{\'e}rio}, {Bonneau}, {Chesneau}, {Clausse}, {Cruzalebes},
  {Perraut}, {Marcotto}, {Roussel}, {Spang}, {McAlister}, {ten Brummelaar},
  {Sturmann}, {Sturmann}, {Turner}, {Farrington}, \& {Goldfinger}}]{Delaa2011}
{Delaa}, O., {Stee}, P., {Meilland}, A., {et~al.} 2011, \aap, 529, A87

\bibitem[{{Domiciano de Souza} {et~al.}(2014){Domiciano de Souza}, {Kervella},
  {Moser Faes}, {Dalla Vedova}, {M{\'e}rand}, {Le Bouquin}, {Espinosa Lara},
  {Rieutord}, {Bendjoya}, {Carciofi}, {Hadjara}, {Millour}, \&
  {Vakili}}]{Domiciano2014}
{Domiciano de Souza}, A., {Kervella}, P., {Moser Faes}, D., {et~al.} 2014,
  \aap, 569, A10

\bibitem[{{Ekstr{\"o}m} {et~al.}(2008){Ekstr{\"o}m}, {Meynet}, {Maeder}, \&
  {Barblan}}]{Ekstrom2008}
{Ekstr{\"o}m}, S., {Meynet}, G., {Maeder}, A., \& {Barblan}, F. 2008, \aap,
  478, 467

\bibitem[{{Fr{\'e}mat} {et~al.}(2005){Fr{\'e}mat}, {Zorec}, {Hubert}, \&
  {Floquet}}]{Fremat2005}
{Fr{\'e}mat}, Y., {Zorec}, J., {Hubert}, A.-M., \& {Floquet}, M. 2005, \aap,
  440, 305

\bibitem[{{Ghosh} {et~al.}(1999){Ghosh}, {Iyengar}, {Ramsey}, \&
  {Austin}}]{Ghosh1999}
{Ghosh}, K., {Iyengar}, K.~V.~K., {Ramsey}, B.~D., \& {Austin}, R.~A. 1999,
  \aj, 118, 1061

\bibitem[{{Gies} {et~al.}(2007){Gies}, {Bagnuolo}, {Baines}, {ten Brummelaar},
  {Farrington}, {Goldfinger}, {Grundstrom}, {Huang}, {McAlister}, {M{\'e}rand},
  {Sturmann}, {Sturmann}, {Touhami}, {Turner}, {Wingert}, {Berger}, {McSwain},
  {Aufdenberg}, {Ridgway}, {Cochran}, {Lester}, {Sterling}, {Bjorkman},
  {Bjorkman}, \& {Koubsk{\'y}}}]{Gies2007}
{Gies}, D.~R., {Bagnuolo}, Jr., W.~G., {Baines}, E.~K., {et~al.} 2007, \apj,
  654, 527

\bibitem[{{Gies} {et~al.}(1998){Gies}, {Bagnuolo}, {Ferrara}, {Kaye},
  {Thaller}, {Penny}, \& {Peters}}]{Gies1998}
{Gies}, D.~R., {Bagnuolo}, Jr., W.~G., {Ferrara}, E.~C., {et~al.} 1998, \apj,
  493, 440

\bibitem[{{Granada} {et~al.}(2013){Granada}, {Ekstr{\"o}m}, {Georgy}, {Krti{\v
  c}ka}, {Owocki}, {Meynet}, \& {Maeder}}]{Granada2013}
{Granada}, A., {Ekstr{\"o}m}, S., {Georgy}, C., {et~al.} 2013, \aap, 553, A25

\bibitem[{{Granada} \& {Haemmerl{\'e}}(2014)}]{Granada2014}
{Granada}, A. \& {Haemmerl{\'e}}, L. 2014, \aap, 570, A18

\bibitem[{{Grundstrom}(2007)}]{Grundstrom2007}
{Grundstrom}, E.~D. 2007, PhD thesis, Georgia State University

\bibitem[{{Grundstrom} \& {Gies}(2006)}]{Grundstrom2006}
{Grundstrom}, E.~D. \& {Gies}, D.~R. 2006, \apjl, 651, L53

\bibitem[{{Hanuschik}(1996)}]{Hanuschik1996}
{Hanuschik}, R.~W. 1996, \aap, 308, 170

\bibitem[{{Hummel} \& {{\v S}tefl}(2001)}]{Hummel2001}
{Hummel}, W. \& {{\v S}tefl}, S. 2001, \aap, 368, 471

\bibitem[{{Hummel} \& {{\v S}tefl}(2003)}]{Hummel2003}
{Hummel}, W. \& {{\v S}tefl}, S. 2003, \aap, 405, 227

\bibitem[{{Ireland} {et~al.}(2006){Ireland}, {Monnier}, \&
  {Thureau}}]{macim2006}
{Ireland}, M.~J., {Monnier}, J.~D., \& {Thureau}, N. 2006, in SPIE Conference
  Series, Vol. 6268

\bibitem[{{Kharchenko} {et~al.}(2005){Kharchenko}, {Piskunov}, {R{\"o}ser},
  {Schilbach}, \& {Scholz}}]{Kharchenko2005}
{Kharchenko}, N.~V., {Piskunov}, A.~E., {R{\"o}ser}, S., {Schilbach}, E., \&
  {Scholz}, R.-D. 2005, \aap, 440, 403

\bibitem[{{Koubsk{\'y}} {et~al.}(2014){Koubsk{\'y}}, {Kotkov{\'a}}, {Kraus},
  {Yang}, {{\v S}lechta}, {Harmanec}, {Wolf}, {Votruba}, {Kub{\'a}t},
  {Kub{\'a}tov{\'a}}, {Niemczura}, \& {{\v S}koda}}]{Koubsky2014}
{Koubsk{\'y}}, P., {Kotkov{\'a}}, L., {Kraus}, M., {et~al.} 2014, \aap, 567,
  A57

\bibitem[{{Koubsk{\'y}} {et~al.}(2012){Koubsk{\'y}}, {Kotkov{\'a}}, {Votruba},
  {{\v S}lechta}, \& {Dvo{\v r}{\'a}kov{\'a}}}]{Koubsky2012}
{Koubsk{\'y}}, P., {Kotkov{\'a}}, L., {Votruba}, V., {{\v S}lechta}, M., \&
  {Dvo{\v r}{\'a}kov{\'a}}, {\v S}. 2012, \aap, 545, A121

\bibitem[{{Kraus} {et~al.}(2012){Kraus}, {Monnier}, {Che}, {Schaefer},
  {Touhami}, {Gies}, {Aufdenberg}, {Baron}, {Thureau}, {ten Brummelaar},
  {McAlister}, {Turner}, {Sturmann}, \& {Sturmann}}]{Kraus2012}
{Kraus}, S., {Monnier}, J.~D., {Che}, X., {et~al.} 2012, \apj, 744, 19

\bibitem[{{Lodieu} {et~al.}(2012){Lodieu}, {Deacon}, {Hambly}, \&
  {Boudreault}}]{Lodieu2012}
{Lodieu}, N., {Deacon}, N.~R., {Hambly}, N.~C., \& {Boudreault}, S. 2012,
  \mnras, 426, 3403

\bibitem[{{Lyubimkov} {et~al.}(2010){Lyubimkov}, {Lambert}, {Rostopchin},
  {Rachkovskaya}, \& {Poklad}}]{Lyubimkov2010}
{Lyubimkov}, L.~S., {Lambert}, D.~L., {Rostopchin}, S.~I., {Rachkovskaya},
  T.~M., \& {Poklad}, D.~B. 2010, \mnras, 402, 1369

\bibitem[{{Makarov}(2006)}]{Makarov2006}
{Makarov}, V.~V. 2006, \aj, 131, 2967

\bibitem[{{McSwain} \& {Gies}(2005)}]{McSwain2005}
{McSwain}, M.~V. \& {Gies}, D.~R. 2005, \apjs, 161, 118

\bibitem[{{Megier} {et~al.}(2009){Megier}, {Strobel}, {Galazutdinov}, \&
  {Kre{\l}owski}}]{Megier2009}
{Megier}, A., {Strobel}, A., {Galazutdinov}, G.~A., \& {Kre{\l}owski}, J. 2009,
  \aap, 507, 833

\bibitem[{{Meilland} {et~al.}(2012){Meilland}, {Millour}, {Kanaan}, {Stee},
  {Petrov}, {Hofmann}, {Natta}, \& {Perraut}}]{Meilland2012}
{Meilland}, A., {Millour}, F., {Kanaan}, S., {et~al.} 2012, \aap, 538, A110

\bibitem[{{Millour} {et~al.}(2011){Millour}, {Meilland}, {Chesneau}, {Stee},
  {Kanaan}, {Petrov}, {Mourard}, \& {Kraus}}]{Millour2011}
{Millour}, F., {Meilland}, A., {Chesneau}, O., {et~al.} 2011, \aap, 526, A107

\bibitem[{{Monnier} {et~al.}(2004){Monnier}, {Berger}, {Millan-Gabet}, \& {ten
  Brummelaar}}]{mirc2004}
{Monnier}, J.~D., {Berger}, J.-P., {Millan-Gabet}, R., \& {ten Brummelaar},
  T.~A. 2004, in New Frontiers in Stellar Interferometry, ed. W.~A. {Traub},
  Vol. 5491, 1370

\bibitem[{{Monnier} {et~al.}(2006){Monnier}, {Berger}, {Millan-Gabet}, {Traub},
  {Schloerb}, {Pedretti}, {Benisty}, {Carleton}, {Haguenauer}, {Kern},
  {Labeye}, {Lacasse}, {Malbet}, {Perraut}, {Pearlman}, \&
  {Zhao}}]{monnier2006}
{Monnier}, J.~D., {Berger}, J.-P., {Millan-Gabet}, R., {et~al.} 2006, \apj,
  647, 444

\bibitem[{{Monnier} {et~al.}(2012){Monnier}, {Che}, {Zhao}, {Ekstr{\"o}m},
  {Maestro}, {Aufdenberg}, {Baron}, {Georgy}, {Kraus}, {McAlister}, {Pedretti},
  {Ridgway}, {Sturmann}, {Sturmann}, {ten Brummelaar}, {Thureau}, {Turner}, \&
  {Tuthill}}]{monnier2012}
{Monnier}, J.~D., {Che}, X., {Zhao}, M., {et~al.} 2012, \apjl, 761, L3

\bibitem[{{Monnier} {et~al.}(2007){Monnier}, {Zhao}, {Pedretti}, {Thureau},
  {Ireland}, {Muirhead}, {Berger}, {Millan-Gabet}, {Van Belle}, {ten
  Brummelaar}, {McAlister}, {Ridgway}, {Turner}, {Sturmann}, {Sturmann}, \&
  {Berger}}]{monnier2007}
{Monnier}, J.~D., {Zhao}, M., {Pedretti}, E., {et~al.} 2007, Science, 317, 342

\bibitem[{{Monnier} {et~al.}(2008){Monnier}, {Zhao}, {Pedretti}, {Thureau},
  {Ireland}, {Muirhead}, {Berger}, {Millan-Gabet}, {Van Belle}, {ten
  Brummelaar}, {McAlister}, {Ridgway}, {Turner}, {Sturmann}, {Sturmann},
  {Berger}, {Tannirkulam}, \& {Blum}}]{mirc}
{Monnier}, J.~D., {Zhao}, M., {Pedretti}, E., {et~al.} 2008, in SPIE Conference
  Series, Vol. 7013

\bibitem[{{Morbey} \& {Brosterhus}(1974)}]{Morbey1974}
{Morbey}, C.~L. \& {Brosterhus}, E.~B. 1974, \pasp, 86, 455

\bibitem[{{Mourard} {et~al.}(2011){Mourard}, {B{\'e}rio}, {Perraut}, {Ligi},
  {Blazit}, {Clausse}, {Nardetto}, {Spang}, {Tallon-Bosc}, {Bonneau},
  {Chesneau}, {Delaa}, {Millour}, {Stee}, {Le Bouquin}, {ten Brummelaar},
  {Farrington}, {Goldfinger}, \& {Monnier}}]{vega2}
{Mourard}, D., {B{\'e}rio}, P., {Perraut}, K., {et~al.} 2011, \aap, 531, A110

\bibitem[{{Mourard} {et~al.}(1989){Mourard}, {Bosc}, {Labeyrie}, {Koechlin}, \&
  {Saha}}]{gcasNature}
{Mourard}, D., {Bosc}, I., {Labeyrie}, A., {Koechlin}, L., \& {Saha}, S. 1989,
  \nat, 342, 520

\bibitem[{{Mourard} {et~al.}(2012){Mourard}, {Challouf}, {Ligi}, {B{\'e}rio},
  {Clausse}, {Gerakis}, {Bourges}, {Nardetto}, {Perraut}, {Tallon-Bosc},
  {McAlister}, {ten Brummelaar}, {Ridgway}, {Sturmann}, {Sturmann}, {Turner},
  {Farrington}, \& {Goldfinger}}]{vegaspie2012}
{Mourard}, D., {Challouf}, M., {Ligi}, R., {et~al.} 2012, in SPIE Conference
  Series, Vol. 8445

\bibitem[{{Mourard} {et~al.}(2009){Mourard}, {Clausse}, {Marcotto}, {Perraut},
  {Tallon-Bosc}, {B{\'e}rio}, {Blazit}, {Bonneau}, {Bosio}, {Bresson},
  {Chesneau}, {Delaa}, {H{\'e}nault}, {Hughes}, {Lagarde}, {Merlin}, {Roussel},
  {Spang}, {Stee}, {Tallon}, {Antonelli}, {Foy}, {Kervella}, {Petrov},
  {Thiebaut}, {Vakili}, {McAlister}, {Ten Brummelaar}, {Sturmann}, {Sturmann},
  {Turner}, {Farrington}, \& {Goldfinger}}]{vega}
{Mourard}, D., {Clausse}, J.~M., {Marcotto}, A., {et~al.} 2009, \aap, 508, 1073

\bibitem[{{Negueruela} \& {Okazaki}(2001)}]{Negueruela2001}
{Negueruela}, I. \& {Okazaki}, A.~T. 2001, \aap, 369, 108

\bibitem[{{Neiner} {et~al.}(2011){Neiner}, {de Batz}, {Cochard}, {Floquet},
  {Mekkas}, \& {Desnoux}}]{Neiner2011}
{Neiner}, C., {de Batz}, B., {Cochard}, F., {et~al.} 2011, \aj, 142, 149

\bibitem[{{Pauls} {et~al.}(2005){Pauls}, {Young}, {Cotton}, \&
  {Monnier}}]{Pauls2005}
{Pauls}, T.~A., {Young}, J.~S., {Cotton}, W.~D., \& {Monnier}, J.~D. 2005,
  \pasp, 117, 1255

\bibitem[{{Pearson} \& {Readhead}(1984)}]{Pearson1984}
{Pearson}, T.~J. \& {Readhead}, A.~C.~S. 1984, \araa, 22, 97

\bibitem[{{Peters} {et~al.}(2013){Peters}, {Pewett}, {Gies}, {Touhami}, \&
  {Grundstrom}}]{Peters2013}
{Peters}, G.~J., {Pewett}, T.~D., {Gies}, D.~R., {Touhami}, Y.~N., \&
  {Grundstrom}, E.~D. 2013, \apj, 765, 2

\bibitem[{{Poeckert}(1981)}]{Poeckert1981}
{Poeckert}, R. 1981, \pasp, 93, 297

\bibitem[{{Pols} {et~al.}(1991){Pols}, {Cote}, {Waters}, \& {Heise}}]{Pols1991}
{Pols}, O.~R., {Cote}, J., {Waters}, L.~B.~F.~M., \& {Heise}, J. 1991, \aap,
  241, 419

\bibitem[{{Quirrenbach} {et~al.}(1997){Quirrenbach}, {Bjorkman}, {Bjorkman},
  {Hummel}, {Buscher}, {Armstrong}, {Mozurkewich}, {Elias}, \&
  {Babler}}]{Quirrenbach1997}
{Quirrenbach}, A., {Bjorkman}, K.~S., {Bjorkman}, J.~E., {et~al.} 1997, \apj,
  479, 477

\bibitem[{{Renard} {et~al.}(2011){Renard}, {Thi{\'e}baut}, \&
  {Malbet}}]{Renard2011}
{Renard}, S., {Thi{\'e}baut}, E., \& {Malbet}, F. 2011, \aap, 533, A64

\bibitem[{{Rivinius} {et~al.}(2013){Rivinius}, {Carciofi}, \&
  {Martayan}}]{Rivinius2013}
{Rivinius}, T., {Carciofi}, A.~C., \& {Martayan}, C. 2013, \aapr, 21, 69

\bibitem[{{Shafter} {et~al.}(1986){Shafter}, {Szkody}, \&
  {Thorstensen}}]{Shafter1986}
{Shafter}, A.~W., {Szkody}, P., \& {Thorstensen}, J.~R. 1986, \apj, 308, 765

\bibitem[{{Silaj} {et~al.}(2014){Silaj}, {Jones}, {Sigut}, \&
  {Tycner}}]{Silaj2014}
{Silaj}, J., {Jones}, C.~E., {Sigut}, T.~A.~A., \& {Tycner}, C. 2014, \apj,
  795, 82

\bibitem[{{Silaj} {et~al.}(2010){Silaj}, {Jones}, {Tycner}, {Sigut}, \&
  {Smith}}]{Silaj2010}
{Silaj}, J., {Jones}, C.~E., {Tycner}, C., {Sigut}, T.~A.~A., \& {Smith}, A.~D.
  2010, \apjs, 187, 228

\bibitem[{{Slettebak}(1966)}]{Slettebak1966}
{Slettebak}, A. 1966, \apj, 145, 121

\bibitem[{{Slettebak}(1982)}]{Slettebak1982}
{Slettebak}, A. 1982, \apjs, 50, 55

\bibitem[{{Smith} {et~al.}(2012){Smith}, {Lopes de Oliveira}, {Motch}, {Henry},
  {Richardson}, {Bjorkman}, {Stee}, {Mourard}, {Monnier}, {Che}, {B{\"u}cke},
  {Pollmann}, {Gies}, {Schaefer}, {ten Brummelaar}, {McAlister}, {Turner},
  {Sturmann}, {Sturmann}, \& {Ridgway}}]{smith}
{Smith}, M.~A., {Lopes de Oliveira}, R., {Motch}, C., {et~al.} 2012, \aap, 540,
  A53

\bibitem[{{Stee} {et~al.}(2012){Stee}, {Delaa}, {Monnier}, {Meilland},
  {Perraut}, {Mourard}, {Che}, {Schaefer}, {Pedretti}, {Smith}, {Lopes de
  Oliveira}, {Motch}, {Henry}, {Richardson}, {Bjorkman}, {B{\"u}cke},
  {Pollmann}, {Zorec}, {Gies}, {ten Brummelaar}, {McAlister}, {Turner},
  {Sturmann}, {Sturmann}, \& {Ridgway}}]{stee}
{Stee}, P., {Delaa}, O., {Monnier}, J.~D., {et~al.} 2012, \aap, 545, A59

\bibitem[{{Tallon-Bosc} {et~al.}(2008){Tallon-Bosc}, {Tallon}, {Thi{\'e}baut},
  {B{\'e}chet}, {Mella}, {Lafrasse}, {Chesneau}, {Domiciano de Souza},
  {Duvert}, {Mourard}, {Petrov}, \& {Vannier}}]{litpro}
{Tallon-Bosc}, I., {Tallon}, M., {Thi{\'e}baut}, E., {et~al.} 2008, in Proc.
  SPIE, Vol. 7013

\bibitem[{{ten Brummelaar} {et~al.}(2005){ten Brummelaar}, {McAlister},
  {Ridgway}, {Bagnuolo}, {Turner}, {Sturmann}, {Sturmann}, {Berger}, {Ogden},
  {Cadman}, {Hartkopf}, {Hopper}, \& {Shure}}]{chara}
{ten Brummelaar}, T.~A., {McAlister}, H.~A., {Ridgway}, S.~T., {et~al.} 2005,
  \apj, 628, 453

\bibitem[{{Thaller} {et~al.}(1995){Thaller}, {Bagnuolo}, {Gies}, \&
  {Penny}}]{Thaller1995}
{Thaller}, M.~L., {Bagnuolo}, Jr., W.~G., {Gies}, D.~R., \& {Penny}, L.~R.
  1995, \apj, 448, 878

\bibitem[{{Thiebaut} \& {Giovannelli}(2010)}]{Thiebaut2010}
{Thiebaut}, E. \& {Giovannelli}, J.-F. 2010, IEEE Signal Processing Magazine,
  27, 97

\bibitem[{{Thom} {et~al.}(1986){Thom}, {Granes}, \& {Vakili}}]{thom}
{Thom}, C., {Granes}, P., \& {Vakili}, F. 1986, \aap, 165, L13

\bibitem[{{Tomasella} {et~al.}(2010){Tomasella}, {Munari}, \&
  {Zwitter}}]{Tomasella2010}
{Tomasella}, L., {Munari}, U., \& {Zwitter}, T. 2010, \aj, 140, 1758

\bibitem[{{Torres} {et~al.}(2010){Torres}, {Andersen}, \&
  {Gim{\'e}nez}}]{Torres2010}
{Torres}, G., {Andersen}, J., \& {Gim{\'e}nez}, A. 2010, \aapr, 18, 67

\bibitem[{{Touhami} {et~al.}(2011){Touhami}, {Gies}, \&
  {Schaefer}}]{Touhami2011}
{Touhami}, Y., {Gies}, D.~R., \& {Schaefer}, G.~H. 2011, \apj, 729, 17

\bibitem[{{Touhami} {et~al.}(2013){Touhami}, {Gies}, {Schaefer}, {McAlister},
  {Ridgway}, {Richardson}, {Matson}, {Grundstrom}, {ten Brummelaar},
  {Goldfinger}, {Sturmann}, {Sturmann}, {Turner}, \&
  {Farrington}}]{Touhami2013}
{Touhami}, Y., {Gies}, D.~R., {Schaefer}, G.~H., {et~al.} 2013, \apj, 768, 128

\bibitem[{{Townsend} {et~al.}(2004){Townsend}, {Owocki}, \&
  {Howarth}}]{Townsend2004}
{Townsend}, R.~H.~D., {Owocki}, S.~P., \& {Howarth}, I.~D. 2004, \mnras, 350,
  189

\bibitem[{{Tycner} {et~al.}(2006){Tycner}, {Gilbreath}, {Zavala}, {Armstrong},
  {Benson}, {Hajian}, {Hutter}, {Jones}, {Pauls}, \& {White}}]{Tycner2006}
{Tycner}, C., {Gilbreath}, G.~C., {Zavala}, R.~T., {et~al.} 2006, \aj, 131,
  2710

\bibitem[{{{\v S}tefl} {et~al.}(2000){{\v S}tefl}, {Hummel}, \&
  {Rivinius}}]{Stefl2000}
{{\v S}tefl}, S., {Hummel}, W., \& {Rivinius}, T. 2000, \aap, 358, 208

\bibitem[{{van Leeuwen}(2007)}]{leeuw2007b}
{van Leeuwen}, F. 2007, \aap, 474, 653

\bibitem[{{Vanbeveren} {et~al.}(1998{\natexlab{a}}){Vanbeveren}, {De Loore}, \&
  {Van Rensbergen}}]{Vanbeveren1998b}
{Vanbeveren}, D., {De Loore}, C., \& {Van Rensbergen}, W. 1998{\natexlab{a}},
  \aapr, 9, 63

\bibitem[{{Vanbeveren} {et~al.}(1998{\natexlab{b}}){Vanbeveren}, {van
  Rensbergen}, \& {De Loore}}]{Vanbeveren1998}
{Vanbeveren}, D., {van Rensbergen}, W., \& {De Loore}, C., eds.
  1998{\natexlab{b}}, Astrophysics and Space Science Library, Vol. 232, {The
  brightest binaries.}

\bibitem[{{Zorec} {et~al.}(2005){Zorec}, {Fr{\'e}mat}, \& {Cidale}}]{Zorec2005}
{Zorec}, J., {Fr{\'e}mat}, Y., \& {Cidale}, L. 2005, \aap, 441, 235

\end{thebibliography}

\begin{appendix}
\section{Visible-image reconstruction process}
\label{sec:reconstruction}

\subsection{Self-calibration}

We used the software MiRA \citep{Thiebaut2010} combined
with an improved self-calibration method \citep{Millour2011} to reconstruct wavelength-dependent images of \object{$\varphi$ Persei} in the visible wavelength range. The data set contains only broadband $V^2$ defined in Table~\ref{tab:spectro} and the wavelength-dependent differential visibilities and phases, but no closure phases. Therefore, the self-calibration method had to be updated: the first step of image-reconstruction is made using the broadband $V^2$ only. The first tentative map is therefore the same for all wavelengths. Then we had to self-calibrate the visibility gains in addition to the differential phases offsets. We outline below the different steps of our method and the additions
made in this work. We also refer to the review of \citet{Pearson1984} for an overview of the self-calibration methods in radio-astronomy. The so-called \emph{SELF-CAL} software and its documentation are available online\footnote{Available at http://self-cal.oca.eu}.

\begin{enumerate}
  \item {\emph{Initial image reconstruction}: we call $I_0(x,y,\lambda_0)$ the first tentative image reconstruction using the $V^2(u,v,\lambda_0)$ only, $\widehat{I_0}(u,v,\lambda_0)$ its Fourier transform, and ${\rm d}V(u,v,\lambda)$, ${\rm d}\phi(u,v,\lambda)$ the differential visibilities and phases provided by VEGA.}

  \item {\emph{Self-calibration}: in the self-calibration process we search for complex gain values $g(u,v)$ that are a function of the $(u,v)$ coordinates and have the following equalities:
\begin{eqnarray}
||\widehat{I_0}(u,v,\lambda)|| & =& ||g(u,v)|| \times {\rm d}V(u,v,\lambda) \\
\arg[\widehat{I_0}(u,v,\lambda)] & = & \arg[g(u,v)] \times {\rm d}\phi(u,v,\lambda)
\end{eqnarray}}

  \item {\emph{$(x,y,\lambda)$ Image reconstruction}: these gain
    values are then applied to the differential phases and
    visibilities, which become amplitude and phase of a complex
    visibility that is stored to a new OIFITS file
    \citep{Pauls2005}. This OIFITS file is fed again to the MiRA
    software to produce a $(x,y,\lambda)$ image.}
\end{enumerate}

This new image serves as a basis to a new step of
self-calibration. The self-calibration process can be repeated until a
reasonable convergence is achieved. Here, we decided to arbitrarily
stop at the second step, as in \citet{Millour2011}.

\subsection{Selecting proper image reconstruction parameters}

For the image reconstruction process, we had to select the field of
view, the pixel size and the reconstruction hyperparameter $\mu$
values.

\paragraph{Field of view}

The field of view of our images was selected thanks to the following
relation:

\begin{equation}
{\rm FOV} = \lambda_{\rm max} / B_{\rm min}
\end{equation}

Here, the smallest available baseline was $B_{\rm min}= 32\,m$ and $
\lambda_{\rm max}=665$\,nm, hence a field of view of 4.2\,mas.

\paragraph{Pixel size}

The pixel size was selected thanks to

\begin{equation}
{\rm \diameter_{\rm pixel}} = \lambda_{\rm min} / B_{\rm max} / 4
.\end{equation}

The factor 4 comes from a double factor 2: the diffraction limit of
the interferometer as defined as in \citet{monnier2007}, and an
additional "super-resolution" factor 2. We have a maximum baseline of
312\,m, which translates into a 0.1\,mas pixel size.

In addition, for computing time considerations, we had to select a
power of 2 for the pixel number. Therefore, we selected images with
64x64 pixels of 0.1\,mas.

\paragraph{Hyperparameter selection}

For the images presented here, we used the total variation
regularization as suggested in \citet{Renard2011}, a number of
iterations per wavelength of 50, and we selected the hyperparameter
value $\mu=10^5$ thanks to the same method as described in
\citet{Domiciano2014}.

\end{appendix}

\end{document}